\documentclass[twocolumn,english,superscriptaddress,notitlepage]{revtex4-1}
\usepackage[T1]{fontenc}
\usepackage[latin9]{inputenc}
\usepackage{graphicx}
\usepackage{amsmath}
\usepackage{float}

\makeatletter
\usepackage{babel}
\usepackage{color}

\makeatother

\mathchardef\mhyphen="2D 

\begin{document}

\title{Counterroating Magnetic Order in the Honeycomb Layers of $\rm NaNi_2BiO_{6-\delta}$}

\author{A. Scheie}
\address{Institute for Quantum Matter and Department of Physics and Astronomy, Johns Hopkins University, Baltimore, MD 21218}

\author{K. Ross}
\address{Department of Physics, Colorado State University, Fort Collins, CO 80523}
\address{Quantum Materials Program, Canadian Institute for Advanced Research (CIFAR), Toronto, Ontario M5G 1Z8, Canada}

\author{P.  Peter  Stavropoulos}
\address{Department of Physics, University of Toronto, Toronto, Ontario, Canada M5S 1A7}

\author{E. Seibel}
\address{Department of Chemistry, Princeton University, Princeton, NJ 08544 }

\author{J. A. Rodriguez-Rivera}
\address{NIST Center for Neutron Research, National Institute of Standards and Technology, Gaithersburg, MD 20899}
\address{Department of Materials Sciences, University of Maryland, College Park, MD 20742}

\author{J. A. Tang}
\address{Department of Chemistry, Johns Hopkins University, Baltimore, MD 21218}

\author{Yi Li}
\address{Institute for Quantum Matter and Department of Physics and Astronomy, Johns Hopkins University, Baltimore, MD 21218}

\author{Hae-Young Kee}
\address{Department of Physics, University of Toronto, Toronto, Ontario, Canada M5S 1A7}
\address{Quantum Materials Program, Canadian Institute for Advanced Research (CIFAR), Toronto, Ontario M5G 1Z8, Canada}

\author{R. J. Cava}
\address{Department of Chemistry, Princeton University, Princeton, NJ 08544 }

\author{C. Broholm}
\address{Institute for Quantum Matter and Department of Physics and Astronomy, Johns Hopkins University, Baltimore, MD 21218}
\address{NIST Center for Neutron Research, National Institute of Standards and Technology, Gaithersburg, MD 20899}
\address{Department of Materials Science and Engineering, Johns Hopkins University, Baltimore, MD 21218}

\date{\today}


\begin{abstract}
We report the magnetic structure and electronic properties of the honeycomb antiferromagnet $\rm NaNi_2BiO_{5.66}$. We find magnetic order with moments along the $c$ axis for temperatures below $T_{c1}=6.3(1)\>{\rm K}$ and then in the honeycomb plane for $T < T_{c2}=4.8(1)\>{\rm K}$ with a counterrotating pattern and an ordering wave vector 
${\bf q}=(\frac{1}{3},\> \frac{1}{3},\> 0.15(1))$. 
Density functional theory and electron spin resonance indicate this is high-spin Ni$^{3+}$ magnetism near a high to low spin transition. The ordering wave vector, in-plane magnetic correlations, missing entropy, spin state, and superexchange pathways are all consistent with bond-dependent Kitaev-$\Gamma$-Heisenberg exchange interactions in  $\rm NaNi_2BiO_{6-\delta}$.
\end{abstract}

\maketitle

\section{Introduction}

The discovery of the exactly solvable Kitaev model with a spin liquid ground state \cite{Kitaev2006} has attracted much attention to the realization and consequences of anisotropic bond dependent exchange interactions on the honeycomb lattice \cite{Takagi2019}. In the last decade various $4d$ and $5d$ electron systems have been found to 
 exhibit Kitaev interactions including $\rm \alpha \mhyphen RuCl_3$ \cite{Plumb2014, Sears2015, Banerjee2016, Banerjee2017} and the iridates \cite{Ganesh2011, Biffin2014, HwanChun2015,Kitagawa2018,Takayama2015,Ruiz2017,Gegenwart2010}. 
However, none of these exhibits the zero field Kitaev spin liquid so  
it would be useful to find more ions which display bond-dependent Kitaev interactions so that the parameter space of materials in which to search for a spin liquid phase can be expanded. In addition, there have been some intriguing predictions of exotic quasiparticles for $S > 1/2$ Kitaev models \cite{Baskaran2008,Koga2018}, but high-spin Kitaev materials are lacking.
Here we present an experimental realization of the magnetic Kitaev-$\Gamma$-Heisenberg \cite{Rau2014} exchange for high-spin Ni$^{3+}$ on a honeycomb lattice, producing the associated conterrotating spiral order.
The resulting magnetism is commensurate in the honeycomb plane and also modulated along the $c$-axis with a wave vector component 0.15(1) $c*$ that is indistinguishable from 1/6 $c*$. Furthermore the magnetism is characterized by strong quantum fluctuations. By demonstrating that $3d$ transition ions such as Ni can exhibit anisotropic bond-dependent exchange, this discovery opens up a whole new class of materials to the search for a Kitaev spin liquid.

\begin{figure}
\centering\includegraphics[scale=0.78]{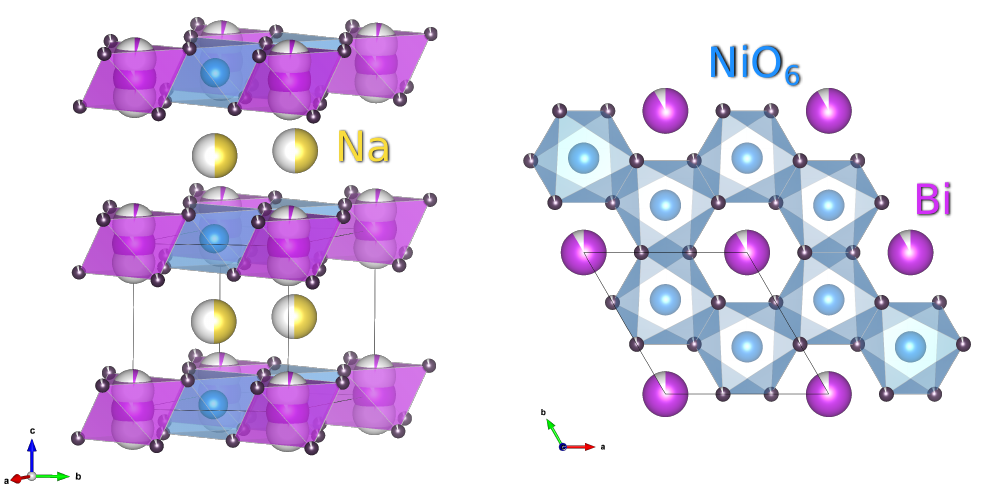}

\caption{Crystal structure of $\rm NaNi_2BiO_{6-\delta}$ from ref. \cite{Seibel2014}.}

\label{flo:CrystalStructure}
\end{figure}

Recently, Seibel et al. discovered and reported  $\rm NaNi_2BiO_{6-\delta}$ which features magnetic Ni ions on a honeycomb lattice \cite{Seibel2014} (Figure \ref{flo:CrystalStructure}). The space group is $P\overline{3}1m$, with lattice parameters $a=b=5.225(3)$ and $c=5.732(5)$ at temperature $T= 2$~K.  Thermogravimetric analysis indicates that $\delta=0.33$ which corresponds to 1/18 oxygen vacancy. 
A Curie-Weiss fit to high temperature susceptibility data yields a Weiss temperature of $\Theta_{CW}=-18.5\>$K and an effective moment of 2.21(1) $\mu_B$/Ni \cite{Seibel2014}. 
Zero-field heat capacity measurements versus $T$ (Fig. \ref{flo:HeatCapacity}) shows two peaks that indicate second-order phase transitions at $T_{c1}=6.3(1)\>{\rm K}$ and $T_{c2}=4.8(1)\>{\rm K}$. The strong magnetic field dependence of these peaks shows these transitions are magnetic in nature. 

Here we report the magnetic structure and properties of $\rm NaNi_2BiO_{6-\delta}$ based on heat capacity, electron spin resonance, density functional theory, and neutron scattering. 
We argue that the counterrotating magnetic order that we have discovered results from  dominant bond-dependent Kitaev exchange within the honeycomb lattices of $\rm NaNi_2BiO_{6-\delta}$, a first example for a Ni based magnet. 

\section{Experiments and Calculations}

\begin{figure}
\centering\includegraphics[scale=0.44]{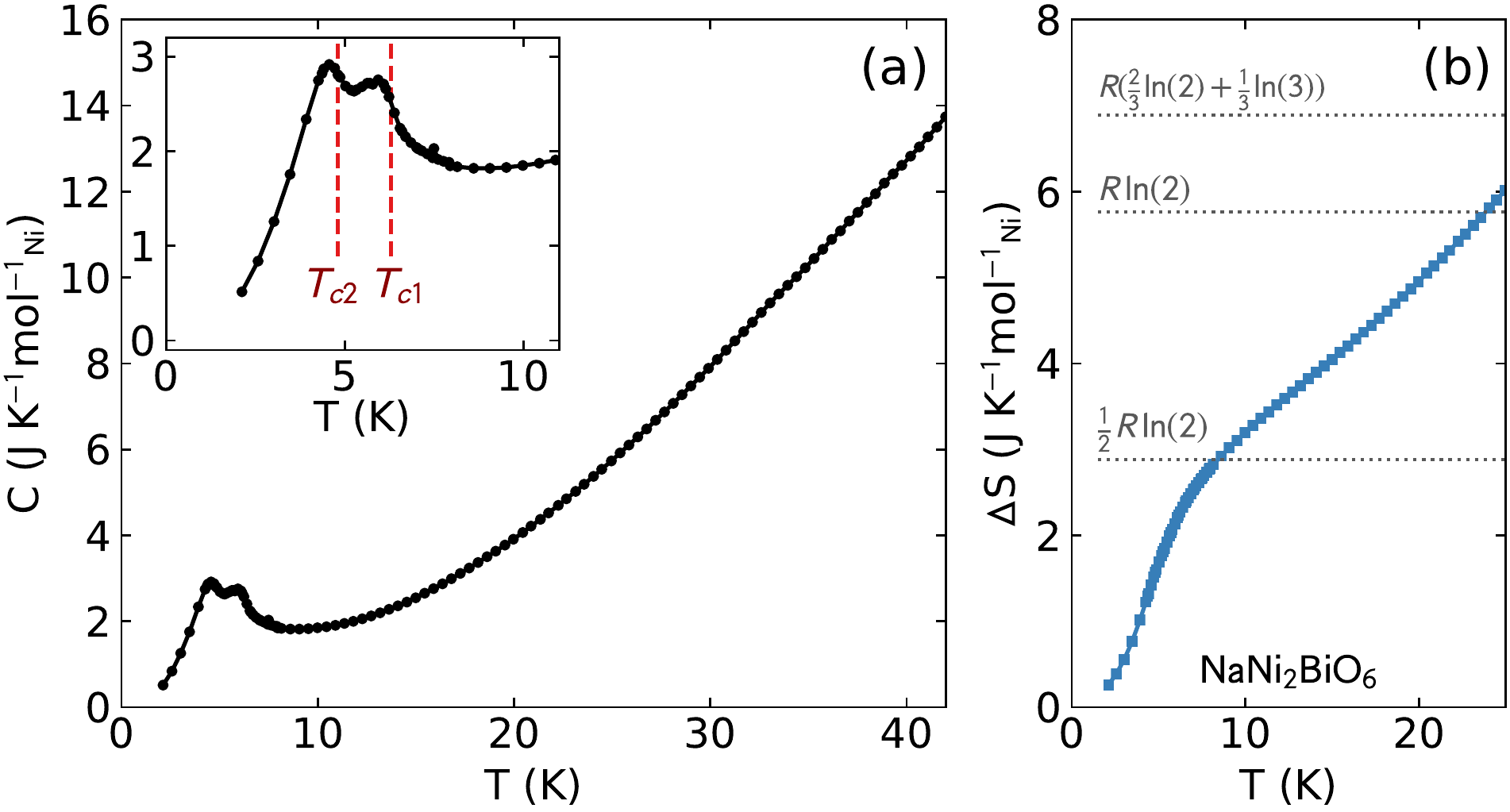}

\caption{Low temperature heat capacity of $\rm NaNi_2BiO_{6-\delta}$. (a) Plot of measured heat capacity. The inset shows two transitions at $T_{c1}=6.3(1)\>{\rm K}$ and $T_{c2}=4.8(1)\>{\rm K}$. (b) Entropy obtained from integrating $C/T$ (extrapolated to zero using a $T^3$ fit).  Note that the lattice contribution to the specific heat has not been subtracted. }

\label{flo:HeatCapacity}
\end{figure}

We measured the heat capacity of $\rm NaNi_2BiO_{6-\delta}$  for $2\>$K$<T<44\>$K using a Quantum Design PPMS \cite{NIST_disclaimer}  (Fig. \ref{flo:HeatCapacity}). Note that the transition temperatures $T_{c1}$ and $T_{c2}$ are associated with the inflection points in heat capacity---see Appendix \ref{flo:HeatCapacityPeak} for details. 
We estimated the overall change in entropy (magnetic and structural) by computing $\Delta S = \int \frac{C_m}{T}dT$ (extrapolating to $C=0$ at $T=0$ using a cubic $T$-dependence).

\begin{figure}
\centering\includegraphics[scale=0.6]{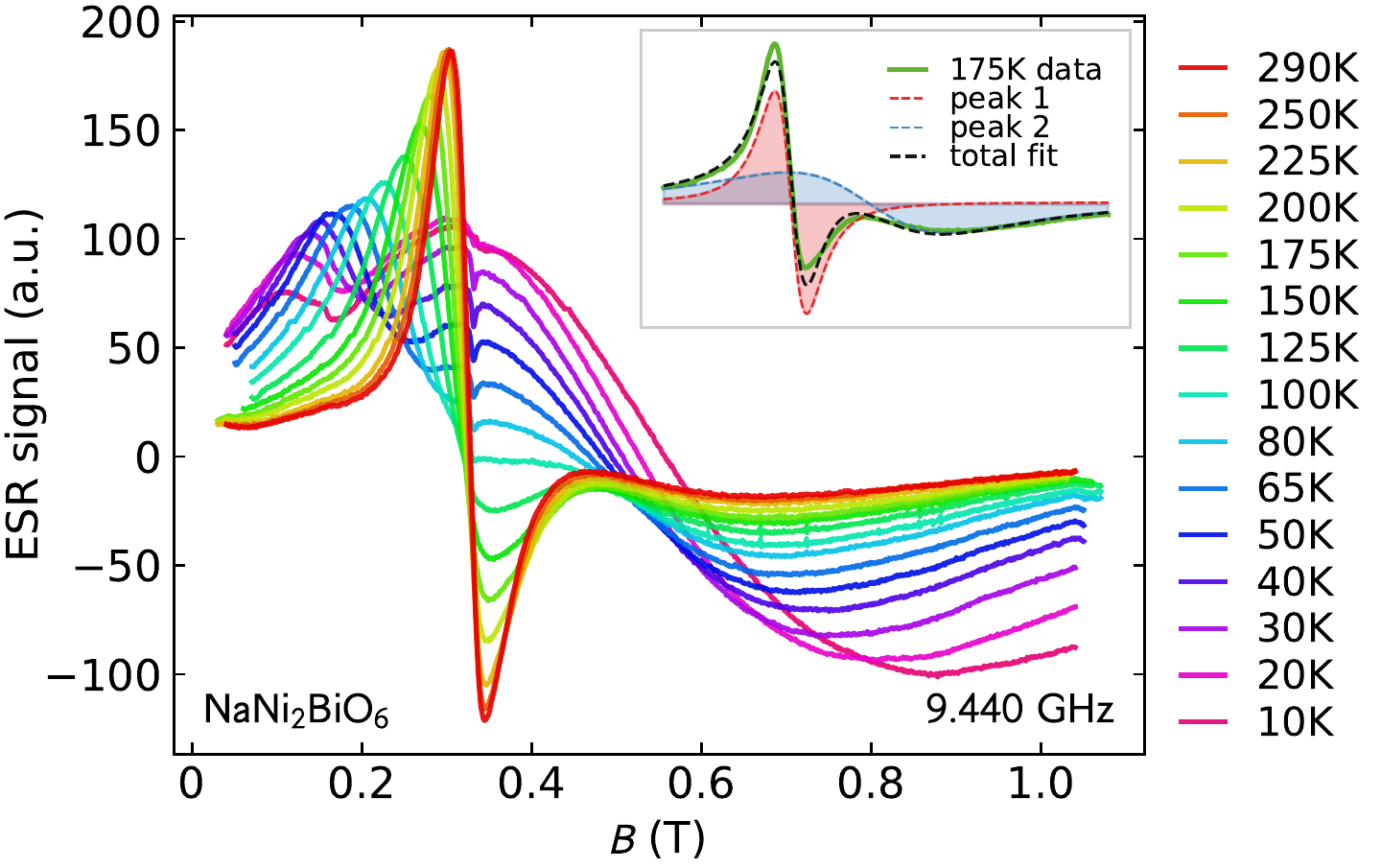}

\caption{ESR data for $\rm NaNi_2BiO_{6-\delta}$, measured between $290\>$K and $10\>$K. The inset showns an example of a two Lorentzian derivative curve fit to the data. Figure \ref{flo:ESR_fits} shows the data extracted from these fits.}

\label{flo:ESR}
\end{figure}

\begin{figure}
\centering\includegraphics[width=0.98\linewidth]{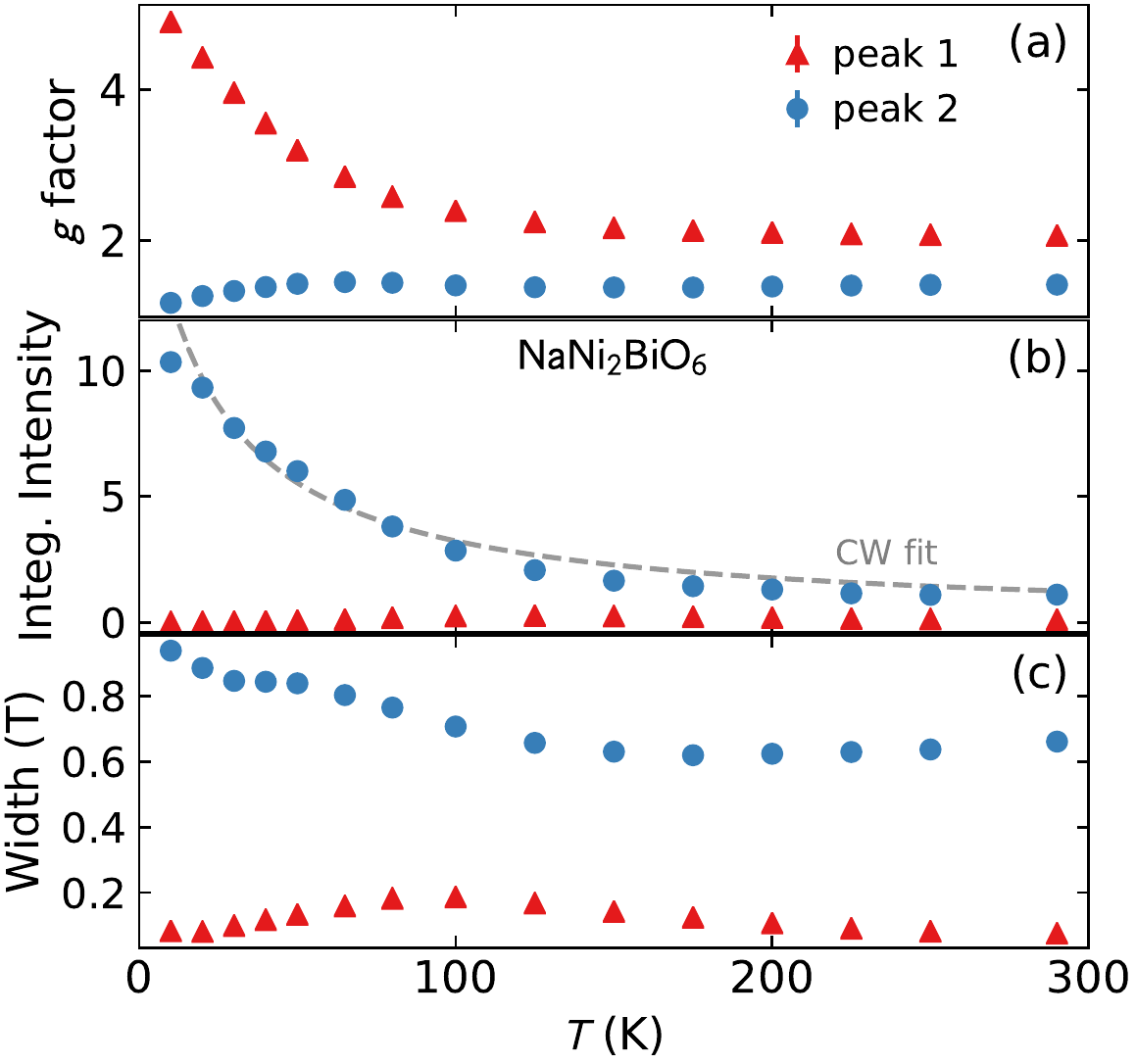}

\caption{Data extracted from two Lorentzian derivative fits to ESR data in Fig. \ref{flo:ESR}. (a) Temperature dependence of the effective $g$ factor, where the red triangles represent the sharp feature (peak 1), and the blue circles represent the broad feature (peak 2). (b) Total integrated intensity of each component. (c) Lorentian FWHM of each resonance. One standard deviation error bars are smaller than the symbol sizes. }

\label{flo:ESR_fits}
\end{figure}

We collected the X-band electron spin resonance (ESR) data shown in Fig. \ref{flo:ESR} on 200 mg of loose powder using a Bruker EMX spectrometer \cite{NIST_disclaimer}. The powder was sealed in a quartz tube filled with argon gas to avoid contact with air. 
Magnetic field scans for temperatures between $10\>$K and $290\>$K were performed at 9.440 GHz, with and without the sample so we can display and analyze difference data that reflect ESR from the sample. Two resonances are visible in the data, so we analyzed the ESR data by fitting to two Lorentzian derivative curves,
 with the results shown in  Fig. \ref{flo:ESR_fits}. There is a small resonance feature at $g=2.0$ (the small jog at 0.33 T in the 100 K to 10 K data), but we did not consider it in our analysis. The lack of temperature dependence and the tiny integrated intensity (0.005(1)\% of the broad resonance) suggests this feature is from contaminants in the sample chamber.
 
To understand the valence state of Ni, we used density functional theory to compute the band-structure and partial density of states (PDOS) of $\rm NaNi_2BiO_{6-\delta}$ and $\rm NaNi_2BiO_{5.66}$, using the OPENMX ab-initio package \cite{Ozaki2003,ozakiopen}. The details of these calculations are discussed in Appendix \ref{app:DFT}.

Finally, 
we performed a neutron scattering experiment on $\rm NaNi_2BiO_{6-\delta}$ using MACS at the NCNR with $4.49\>$g loose powder of anhydrous $\rm NaNi_2BiO_{6-\delta}$ loaded in a sealed aluminum can under 1 atm helium at room temperature. (Multiple attempts have failed to produce sizeable single crystals of this material.) The monochromator was set to double focusing with a pre-monochromator aperture of 360 mm x 360 mm. The data are shown in Fig. \ref{flo:Neutronscans}. We measured the momentum ($Q$) dependence of elastic ($E_i = E_f=5 \>{\rm meV}$, $\hbar\omega =0$) and inelastic ($E_i=4.1 \>{\rm meV}$, $E_f=3.7 \>{\rm meV}$, $\hbar\omega = 0.4$) scattering for temperatures between $1.8\>$K and $20\>$K. We also measured the full excitation spectrum at $T=1.8\>$K (below both heat capacity peaks), $T=5\>$K (in between the heat capacity peaks), and at $T=10\>$K (above both heat capacity peaks).  We converted the ratio of detector to monitor count rates to absolute values of the partial differential scattering cross section
\begin{equation}
\frac{d^2 \sigma}{d\Omega d E_0}= N \frac{k_f}{k_i} \big(\frac{\gamma r_0}{2}gf({\bf Q})\big)^2 2 {\cal S}({\bf Q},\omega),
\end{equation}
by normalizing to the (001) nuclear Bragg peak in accord with Ref. \cite{AbsoluteUnits}. Here $\gamma r_0 = 0.5390 \times 10^{-12}$ cm, $g \approx 2$ is the g-factor for Ni, $f(Q)$ is the magnetic form factor for Ni \cite{formfactors} and ${\cal S}({\bf Q},\omega)$ is the spherically-averaged dynamic correlation function.  
Empty can measurements were subtracted from the data presented in Fig. \ref{flo:Neutronscans}(a) and Fig. \ref{flo:Neutronscans}(c)-(e) with a self-shielding factor of 0.93. The horizontal line of diminished intensity at $\hbar \omega = 1.3\>{\rm meV}$ in panels (c)-(e) is is associated with removal of the incident beam beryllium filter for $E_i > 5 \>{\rm meV}$ ($\hbar \omega > 1.3 \>{\rm meV}$). This causes a slight offset in intensity for a small range of $E_i$ near the filter edge that is probably related to higher order Bragg diffracted neutrons that reach the sample when the Be filter is removed and then transfer $\hbar\omega_{\lambda/2} =4E_i-E_f=4\hbar\omega+3E_f$ to the sample in a high energy inelastic scattering process.

\section{Results and Analysis}

\subsection{Heat Capacity and Entropy}

Bearing in mind that we do not separate magnetic and lattice based entropy here,  the
heat capacity data reveals much less entropy recovered across the phase transitions than one would expect for complete magnetic order. If we assume that the oxygen deficiency produces a 2:1 mixture of low-spin Ni$^{3+}$ ($S=1/2$) and Ni$^{2+}$ ($S=1$) (as suggested in ref. \cite{Seibel2014}), the total magnetic entropy would be $\Delta S=R (2/3\ln(2) +1/3 \ln(3))$. However, the entropy recovered between $2\>$K and $10\>$K is only 41\% of this entropy [see Fig. \ref{flo:HeatCapacity}(b)]. 
As we shall show below, the actual orbital configuration of Ni is intermediate between $S=3/2$ and $J=1/2$. This suggests entropy between $R \ln (4)$ and $R \ln (2)$ 
---which makes the discrepancy with the measured change in entropy across the phase transition even larger.
Such missing entropy is common in quasi-2D materials due to short-range 2D correlations developing at higher temperatures \cite{Regnault1990,Nair2018}.
Unfortunately no non-magnetic analogue to $\rm NaNi_2BiO_{6-\delta}$ is available, so we are unable to determine how much additional magnetic entropy is recovered at higher temperatures (see Appendix \ref{app:PhononSubtraction} for details). Nonetheless, it is clear that the change in entropy across the second order phase transitions is significantly less than the full entropy of a local moment per site. 

It was recently theoretically shown that the high-spin Kitaev model has a a finite $T$ entropy plateau upon cooling \cite{Oitmaa2018,Koga2018}. 
The phenomenon of missing entropy is seen in the Ni$^{2+}$ honeycomb compounds $\rm Na_3Ni_3SbO_6$ and $\rm Li_3Ni_3SbO_6$ \cite{Zvereva2015}, consistent with the predicted $\frac{1}{2} R \ln (3)$ incipient entropy plateau of the $S=1$ Kitaev model \cite{Koga2018,Oitmaa2018}.  For the $S=3/2$ Kitaev model, the expected entropy plateau is at $\frac{1}{2} R \ln(2)$ with bond anisotropy and $\frac{1}{2} R \ln(4)$ for the isotropic case \cite{Oitmaa2018}. The entropy recovered over the transition in $\rm NaNi_2BiO_{6-\delta}$ is close to $\frac{1}{2} R \ln(2)$, the value predicted for the $J=1/2$ Kitaev model. (The plateau is smeared out at least partly due to phonon specific heat.) The precise value notwithstanding, reduced change in entropy associated with the phase transitions in $\rm NaNi_2BiO_{6-\delta}$ is consistent with quasi-2D order and correlations at higher temperatures, possibly the incipient entropy plateau of the Kitaev model.

\subsection{Electron Spin Resonance and Density Functional Theory}

To examine the origins of magnetism in $\rm NaNi_2BiO_{6-\delta}$ we use electron spin resonance, which provides information about the nature and anisotropy of local moments in insulating solids. 
Figures \ref{flo:ESR} and \ref{flo:ESR_fits} show the high temperature X-band ESR spectrum, which has a sharp resonance at $g=2.07$ and a broad resonance at $g=1.42$. 
Upon cooling, the sharp resonance looses spectral weight and shifts to lower field (higher effective $g$-factor) while the broad resonance grows stronger and shifts to higher fields (lower effective $g$-factor). 
The overall signal intensity follows a Curie-Weiss law [Fig. \ref{flo:ESR_fits}(b)] consistent with typical transition ion behavior \cite{AbragamBleaney}. A fit to the ESR intensity data above $20\>$K yields $\Theta_{CW}=-20(4)$ K, in agreement with magnetic susceptibility measurements. 

Generally, broad resonances are associated with high-spin ($S > 1/2$) ions that are subject to crystal field splitting while sharper resonances are associated with pure $S=1/2$ ions \cite{AbragamBleaney}. The effective $g$-factors of the two resonances are consistent with this: The broad ESR resonance has an effective $g$ factor of $g=1.42$, suggesting a high-spin state. Meanwhile, the sharp resonance has an effective $g=2.07$, consistent with $S=1/2$ magnetism. The puzzle is reconciling this with the stoichiometry and structure of $\rm NaNi_2BiO_{6-\delta}$. 
It was originally suggested that $\rm NaNi_2BiO_{6-\delta}$ has 2/3 $S=1/2$ (Ni$^{3+}$) and 1/3 $S=1$ (Ni$^{2+}$) \cite{Seibel2014}. Naively therefore, one might associate the broad resonance with the $S=1$ Ni$^{2+}$ sites and the sharp resonance with $S=1/2$ Ni$^{3+}$ sites. However, the sharp resonance carries only 10-15\% of the spectral weight at high temperatures, which does not square with the majority spins being $S=1/2$. 
Even more puzzling is the fact that the sharp $S=1/2$ resonance nearly vanishes at low temperatures. This suggests some kind of thermal depopulation and is very difficult to reconcile with a fixed ratio of  Ni$^{2+}$ and Ni$^{3+}$ set by the oxygen content. To understand these two ESR resonances we turn to density functional theory.

\begin{figure}
\centering\includegraphics[scale=0.46]{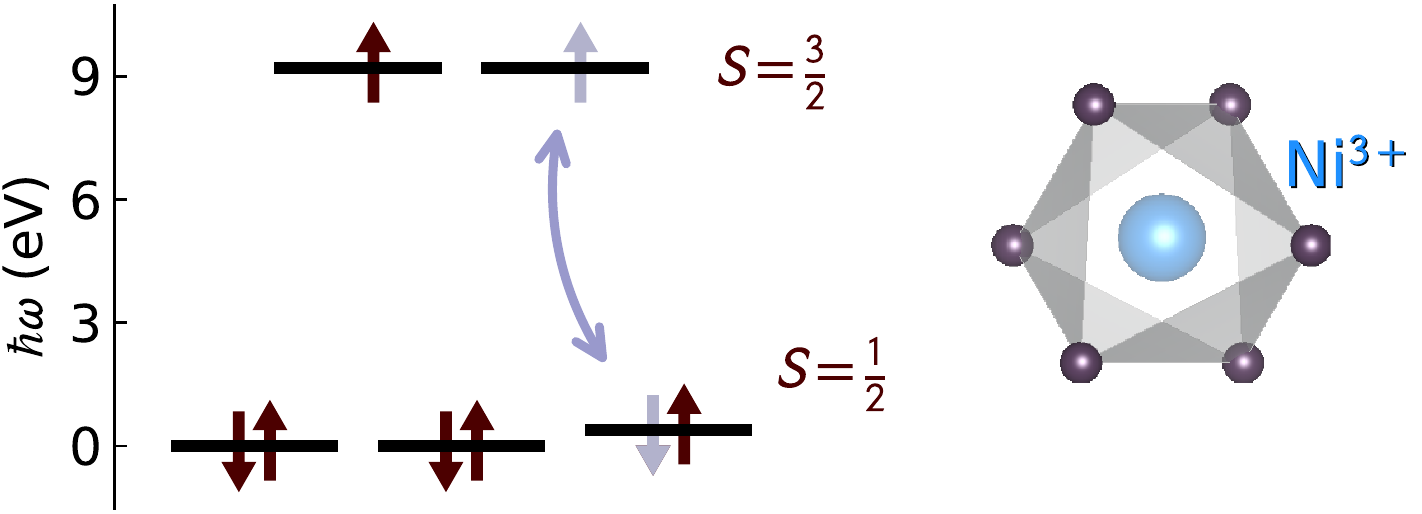}

\caption{Electron orbital energies for Ni$^{3+}$ in $\rm NaNi_2BiO_{6-\delta}$, calculated based on the point charge model using PyCrystalField \cite{PyCrystalField}. The level splitting with a full octahedron has $S=1/2$ in the low-spin case or $S=3/2$ in the high-spin case. SOC is neglected.}

\label{flo:LevelSplitting}
\end{figure}

\subsubsection*{Density Functional Theory}
We found Bi $s$ and O $p$-orbitals form covalent bonds (see the partial density of states (PDOS) in Appendix \ref{app:DFT} Figs. \ref{flo:PDOS1} and \ref{flo:PDOS2}). For 1/18 missing oxygens in $\rm NaNi_2BiO_{5.66}$, electron charge is redistributed between Bi and O so as to quarter fill the Ni $e_g$-orbitals and fill the $t_{2g}$-orbitals.  In other words, all Ni ions are trivalent Ni$^{3+}$ with the $3d^7$ electron configuration. This behavior is independent of the strength of spin-orbit coupling. Thus, we propose that all Ni ions are Ni$^{3+}$ and not a mixture of Ni$^{3+}$ and Ni$^{2+}$ as previously proposed \cite{Seibel2014}.
When Hubbard $U$ and Hund's coupling are included in LDA+SOC+U, the systems develops a local moment for any finite U, indicating that Hund's coupling is strong enough to favor the high spin state $S=3/2$ (Figure \ref{flo:LevelSplitting}).



A natural way to produce a thermally depopulating sharp $S=1/2$ ESR resonance is if the Ni$^{3+}$ high-spin and low-spin states are close in energy (see Fig. \ref{flo:LevelSplitting}). If the $S=3/2$ state is $\sim 10$ meV lower in energy than the $S=1/2$ state, the Ni$^{3+}$ ions would have equally populated $S=3/2$ and $S=1/2$ at 300 K (with a ESR spectrum ratio of $sharp/total = 16.6$\%). For temperatures below 100 K, however, the sharper $S=1/2$ resonance would shrink and the broad $S=3/2$ resonance would grow with the typical Curie-Weiss behavior. This is precisely what we observe.

To test this hypothesis, we compare to experimental quantities: the effective moment from susceptibility $\mu = 2.21(1) \> \mu_B$/Ni, the relative weights of the ESR signals (ESR signal is proportional to $S(S+1)$, see eq. 2.55 in ref. \cite{AbragamBleaney}), and allowing for thermal depopulation of one of the resonances. The results are in Table \ref{tab:SpinStates}, which clearly favors the high-spin Ni$^{3+}$  hypothesis. 

\begin{table}[]
    \centering
    \caption{Predicted effective moment and relative weight of the sharp ESR signal for a 2:1 mixture of $S=1/2$ Ni$^{3+}$ and $S=1$ Ni$^{2+}$ ions and uniform $S=3/2$ Ni$^{3+}$ with thermally populated $S=1/2$. Effective moment is calculated using the $g$ factors from ESR measurements. The experimental values are on the right, and they agree best with a uniform Ni$^{3+}$ state.}
    \begin{ruledtabular}
    \begin{tabular}{c|c c |c }
        Ni$^{3+}$ & $S=\frac{1}{2}$  & $S=\frac{1}{2},\frac{3}{2}$  & Exp.\\
        Ni$^{2+}$ &  $S=1$  &  $none$   & \\
         \hline
        $\mu_{eff}$ ($\mu_B$) & 1.829 & 2.298 & 2.21(1)\\
        ESR $\rm \frac{sharp}{total}$ & 43\% & 16.6\% & 13(2)\%
    \end{tabular}
    \end{ruledtabular}
    \label{tab:SpinStates}
\end{table}


The situation is complicated by the presence of spin orbit coupling.
The spin ($S=3/2$) and orbital ($L=1$) angular momentum  states of octahedrally coordinated Ni$^{3+}$  $3d^7$ are subject to atomic spin-orbit coupling (SOC $\lambda_{Ni3+}=34$ meV \cite{AbragamBleaney}) enhanced by covalent bonding with the Bi ions (see section \ref{sec:Discussion} and ref. \cite{stavropoulos2019microscopic}). This can lead to an effective $J=1/2$ singlet at low temperatures \cite{Cobalt_Liu2018}. 
To examine this, we computed a PDOS using density functional theory including single-ion Ni$^{3+}$ spin orbit coupling and a trigonal distortion of the oxygen octahedra. These calculated results indicate an intermediate state between $J=1/2$ and $S=3/2$ due to the interplay between  trigonal distortion and SOC. This intermediate state is in-between the limit $H_{trigonal} >> H_{SOC}$ where the $S$ basis is valid and the limit $H_{trigonal} << H_{SOC}$ where the $J$ basis is valid, making the ground state eigenket not easily expressible in either form.
Computing $\langle J_z \rangle$ for a single ion using a Kanamori Hamiltonian gives values between 0.8 and 1.2, depending on SOC---neither 3/2 nor 1/2 (see Appendix \ref{app:localmoment}).
Thus, the ground state is not simply $S=\frac{3}{2}$ but a mixed $S=3/2$, $J=1/2$ state. This may explain the unusual temperature-dependent $g$-factor for the broad resonance.



In the end, the ESR data combined with DFT calculations are evidence for uniform Ni$^{3+}$ with a mixed $S=3/2$, $J=1/2$ state.  Our observation through ESR of thermal depopulation of the low spin state in favor of the high spin state conforms with their energetic proximity: Ni$^{3+}$ has previously been found both in the low spin  \cite{Stoyanova1994,sanz2011optical,Meskine2005} and in the high spin state \cite{MOHANRAM1983,Reinen1974}, depending upon the ligand environment. Significantly, the  high-spin Ni$^{3+}$ and orbital coupling to Bi paves the way for bond-dependent anisotropic interactions, as we shall explain below.

\subsection{Neutron Scattering}

\begin{figure}
\centering\includegraphics[width=0.99\linewidth]{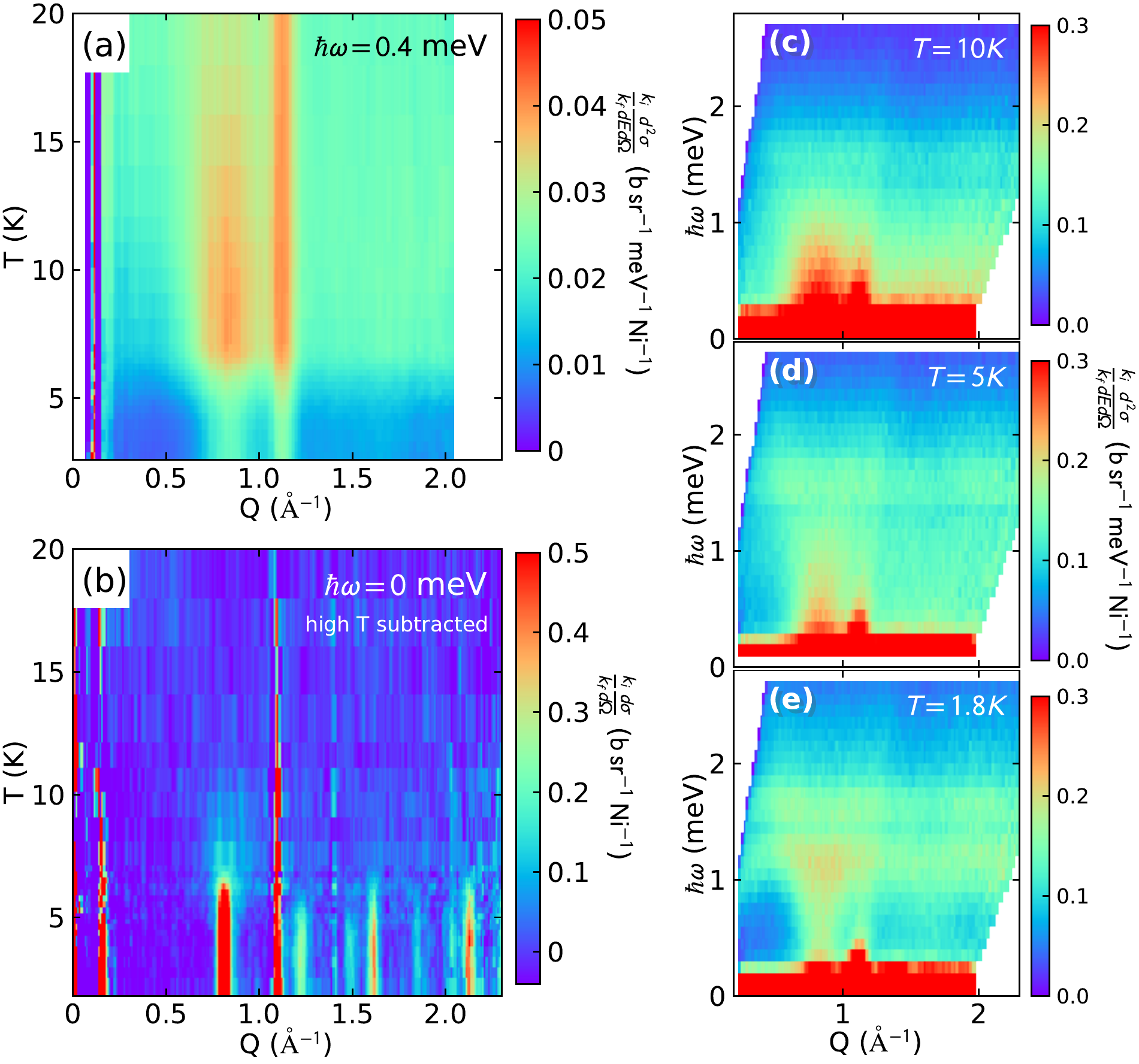}

\caption{Neutron scattering cross section for $\rm NaNi_2BiO_{6-\delta}$. (a) Inelastic temperature scan at $E_i=4.1 \>{\rm meV}$, $E_f=3.7 \>{\rm meV}$ ($\hbar \omega =  E_f - E_i = 0.4 \>{\rm meV}$). (b) Elastic temperature scan ($E_i = E_f=5.0 \>{\rm meV}$, $\hbar \omega=0$), revealing magnetic Bragg peaks that emerge below the transition temperatures. High $T$ data were subtracted to isolate $T$-dependent magnetic diffraction.  The strong feature at 1.1 \AA$^{-1}$  is remnants of a subtracted nuclear Bragg peak. (c-e) Inelastic neutron scattering data at $1.8\>$K, $5\>$K, and $10\>$K.}

\label{flo:Neutronscans}
\end{figure}

The temperature-dependent elastic neutron scattering data in Fig. \ref{flo:Neutronscans}(b) show new Bragg peaks appearing at low temperatures. The onset temperature matches $T_{c1}$ and $T_{c2}$ determined from heat capacity data, indicating that these anomalies mark magnetic phase transitions. The inelastic $\hbar \omega =0.4 \>{\rm meV}$ scattering data in panel (a) show an increase in paramagnetic diffuse scattering for $T > T_{c1},T_{c2}$, and in particular for wave vector transfer $Q$ near the 0.81 \AA$^{-1}$ magnetic peak. The integrated intensity of this inelastic peak, shown versus temperature in Fig. \ref{flo:OrderParameter}(d), is highest at $7\>$K, and then gradually diminishes upon warming.

While it may look like the intensity of the peak in inelastic scattering near the 1.1 \AA$^{-1}$ nuclear Bragg peak in Fig. \ref{flo:Neutronscans}(a) is enhanced above the transition, Gaussian fits to the $Q$-dependent intensity at  each temperature show the integrated intensity of the peak is independent of temperature near $T_c$. The apparent temperature dependence is actually in a $Q$-independent diffuse background that presumably then has a magnetic origin. 

The fixed temperature full-spectrum scans in Fig. \ref{flo:Neutronscans}(c)-(e) provide more information about the the magnetic excitations. The data in Fig. \ref{flo:Neutronscans}(e) resemble powder-averaged inelastic scattering from spin waves with a bandwidth $\approx 2\>$meV, which is the bandwidth estimated from the Curie-Weiss temperature: $\frac {3 k_B}{(S+1)} \Theta_{CW} = 1.91\>$meV for $S=3/2$. 
(A derivation of this equation, which does not deal with the mixed S=3/2, J=1/2 state case, is given in Appendix \ref{app:BandwidthToCW}.) The spin-wave-like excitations and the appearance of low temperature Bragg peaks show the transitions around $5 \>$K are to long-range ordered magnetism.


The $10 \>$K data in Fig. \ref{flo:Neutronscans}(c) shows that spin correlations persist at temperatures well above the upper phase transition. This is consistent with expectations for a frustrated quasi-two-dimensional magnet and with an incipient entropy plateau above $T_{c1}$. 

The dynamic magnetic moment can be computed from the inelastic spectral weight per formula unit using 
\begin{equation}
\langle m^2 \rangle = \frac{3 \mu_B^2 \iint   (1+e^{- E/k_B T}) [ {\cal S}(Q,E)] Q^2 dQ dE}{ \int Q^2 dQ }
\end{equation}
integrated from 0.3 meV to 2.5 meV and from 0.5 \AA$^{-1}$ to 1.9 \AA$^{-1}$, where detailed balance has been employed. 
We find $\langle m^2 \rangle=3.3(7)$ $\rm \mu_B^2$/Ni ion at 1.8 K, 3.6(7) $\rm \mu_B^2$/Ni at 5 K, and 4.1(8) $\rm \mu_B^2$/Ni at 10 K. (Comparison to total moment estimates is made below.) 
These values ought to be taken cautiously because inelastic spectral weight from phonons was not excluded from  the integrals.
That being said, the phonon scattering at $1.8 \>$K and at low $Q$ is relatively weak (phonon intensity varies as $\propto Q^2$), making in particular the result at $1.8 \>$K reliable. 

\subsubsection*{Magnetic Structure:}

\begin{figure}
\centering\includegraphics[scale=0.45]{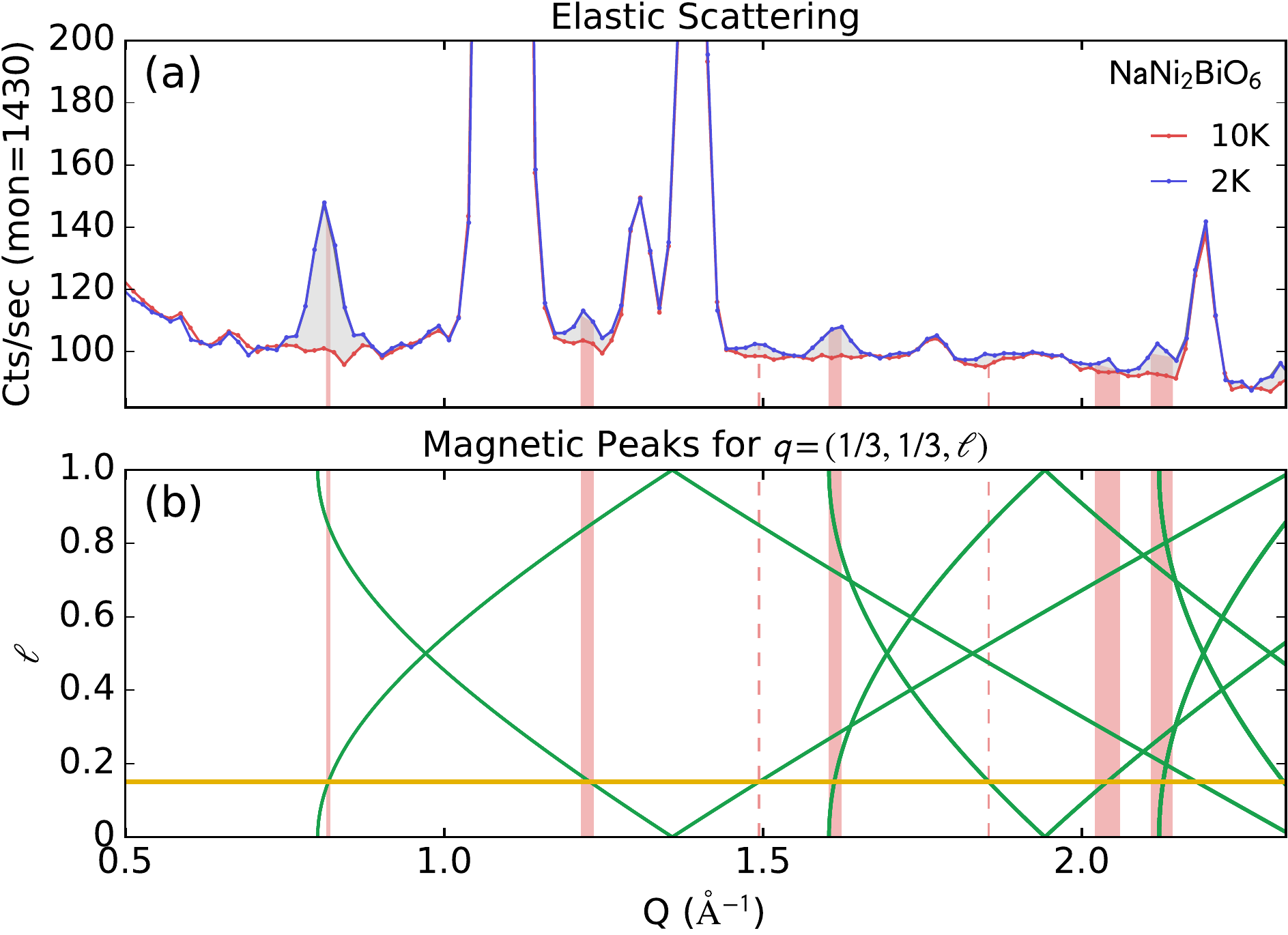}

\caption{Ordering wave vector of $\rm NaNi_2BiO_{6-\delta}$. (a) Plot of elastic neutron scattering at 2~K and 10~K, showing the appearance of additional Bragg peaks at low temperatures. The five strongest temperature-dependent Bragg peaks are indicated with pink vertical bars. (b) Plot of theoretically predicted magnetic peaks (green lines) from ${\bf Q}_i = {\bf \tau}_i \pm {\bf q}$ with $q=(1/3, \>1/3, \> \ell)$, where $\ell$ varies along the $y$ axis. The horizontal yellow line shows $\ell = 0.154$ which correctly indexes the observed Bragg peaks. The vertical pink dashed lines show smaller Bragg peaks also indexed by $q=(1/3, \>1/3, \> 0.154)$.}

\label{flo:OrderingVector}
\end{figure}

Using the elastic scattering data from $\rm NaNi_2BiO_{6-\delta}$, we can determine the low $T$ magnetic structure.
The first step is to identify the wave vector characterizing the magnetic order. 
We compared the wave vectors of the five strongest temperature-dependent Bragg peaks to those calculated from $|{\bf Q}_i| = |{\bf G}_i \pm {\bf q}|$. Here ${\bf G}_i$ are nuclear Bragg peaks and ${\bf q}$ is a symmetry-allowed ordering wave vector in the $P \bar 31m$ space group \cite{Kovalev}. The error bars in experimental peak locations (represented visually by the widths of the vertical bands in Fig. \ref{flo:OrderingVector}) were determined from the range of fitted Gaussian peak locations for elastic data at temperatures below 4K. Visual comparisons, as in Fig. \ref{flo:OrderingVector}(b), allowed us to identify the correct magnetic wave vector $\bf q$.  The only symmetry allowed ordering wave vector that can account for the five strongest magnetic Bragg peaks is ${\bf q} = (\frac{1}{3},\> \frac{1}{3},\> 0.15(1))$. 
As Fig. \ref{flo:OrderingVector} shows, this ordering wave vector also correctly indexes weaker magnetic Bragg peaks at 1.49 \AA$^{-1}$ and 1.85 \AA$^{-1}$. This wave vector means the magnetic unit cell encompasses three nuclear unit cells in the $ab$ plane, and has a characteristic wave length of $c/0.154(11) =6.5(5) c = 37(3)$ \AA~along the $c$ axis. While the $c$-component of the magnetic wave vector could be incommensurate, it is experimentally indistinguishable from the commensurate value of 1/6.

\begin{table*}
\caption{Irreducible Representations and associated basis vectors (BVs) for space
group $P \bar 31m$ and propagation vector $q=(\frac{1}{3},\frac{1}{3},0.154)$. The $\chi^{2}$ values gauge the quality of a Rietveld refinement
to the 1.8K data using the FullProf suite. Refinements to the 1.8~K data cannot distinguish between $\Gamma_1$ and $\Gamma_2$, but symmetry considerations preclude $\Gamma_1$ and $\Gamma_3$ leaving $\Gamma_2$ as the only option consistent with the data.}
\begin{ruledtabular}%
\begin{tabular}{ccccc|cc|cc|cc}
IRs & $\psi_{\nu}$ & component & Ni1 & Ni2 & $\chi^{2}$(5~K) & BVs & $\chi^{2}$(1.8~K) & BVs & $\chi^{2}$(1.8~K) & BVs \tabularnewline
\hline 
$\Gamma_{1}$  & $\psi_{1}$ & Real & (1.5 0 0) & (0 -1.5 0) & 14.1 & &  &  & 9.6 & 0.202  \tabularnewline
 &  & Imaginary & ($-\frac{\sqrt{3}}{2}$ $-\sqrt{3}$ 0) & ($\sqrt{3}$ $\frac{\sqrt{3}}{2}$ 0) &  &  &  &  &  &  \tabularnewline
$\Gamma_{2}$  & $\psi_{2}$ & Real & (1.5 0 0) & (0 1.5 0) & 14.1 & & 9.7 & 0.183 &  &  \tabularnewline
 &  & Imaginary & ($-\frac{\sqrt{3}}{2}$ -$\sqrt{3}$ 0) & (-$\sqrt{3}$ $-\frac{\sqrt{3}}{2}$ 0) &  &  &  &  &  &  \tabularnewline
$\Gamma_{3}$  & $\psi_{3}$ & Real & (1.5 0 0) & (0 -1.5 0) & 5.8 & 0.0 &  &  &  &   \tabularnewline
 &  & Imaginary & ($\frac{\sqrt{3}}{2}$ $\sqrt{3}$ 0) & (-$\sqrt{3}$ $-\frac{\sqrt{3}}{2}$ 0) &  &  &  &  &  & \tabularnewline
 & $\psi_{4}$ & Real & (0 0 3) & (0 0 -3) &  & 0.314 &  & 0.366 &  & 0.337  \tabularnewline
 &  & Imaginary & (0 0 0) & (0 0 0) &  &  &  &  &  &  \tabularnewline
\end{tabular}\end{ruledtabular}
\label{flo:IrrRepTable}

\end{table*}

\begin{figure*}
\centering\includegraphics[scale=0.6]{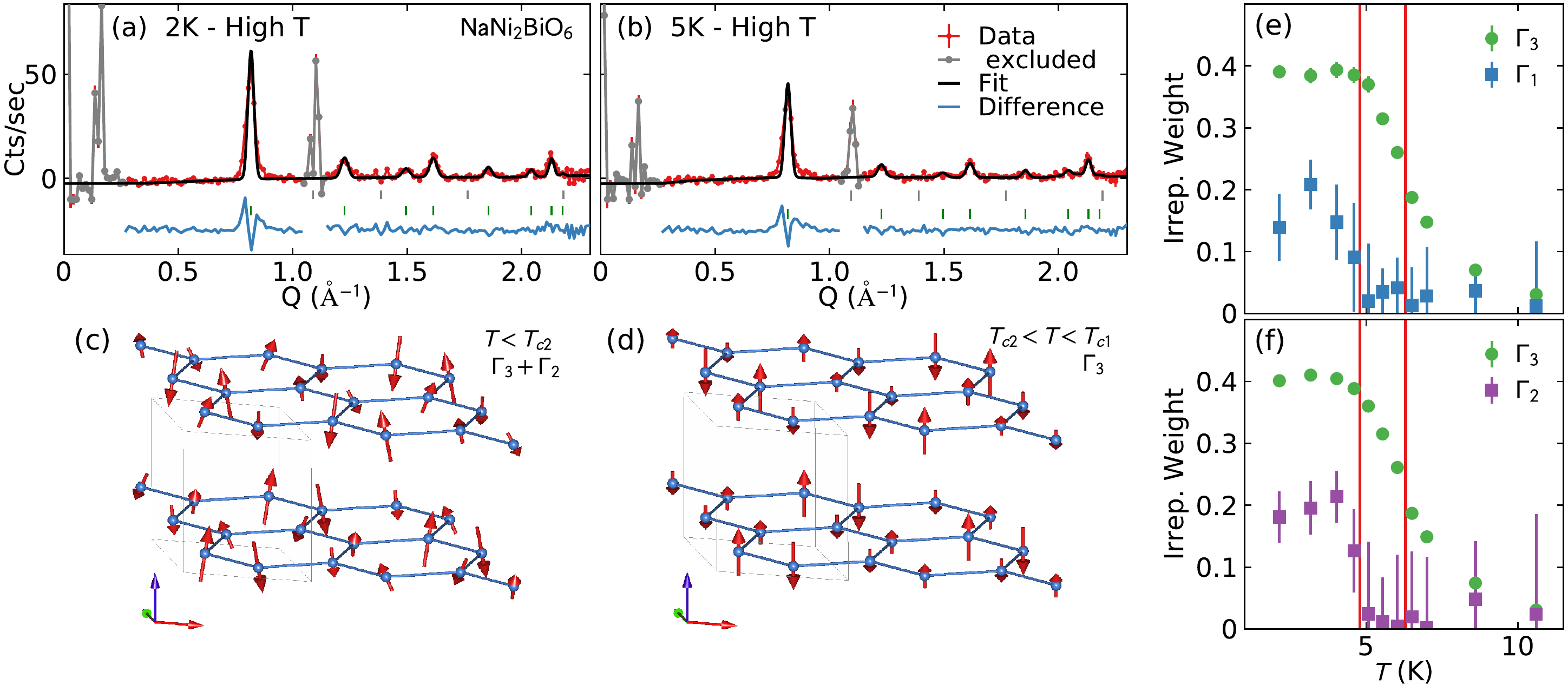}

\caption{Magnetic refinement of $\rm NaNi_2BiO_{6-\delta}$. (a) Refinement of $2\>$K elastic neutron scattering data with high temperature data ($12\>$K-$20\>$K) subtracted. The slight increase with Q in the background difference intensity level away from magnetic Bragg peaks can result from the change in the Debye Waller factor with $T$. This produces a $Q^2$ dependence of the difference intensity for low $Q$. The resulting magnetic structure, shown in panel (c), has an in-plane component to the spins ($\Gamma_2+\Gamma_3$). (b) Refinement of $5\>$K neutron data with high temperature data subtracted. The magnetic structure, shown in panel (d), has all spins aligned along the $c$ axis ($\Gamma_3$). Panels (e) and (f) show the temperature dependence of refined irrep weights of $\Gamma_3$ in combination with $\Gamma_1$ and $\Gamma_2$, respectively. The red vertical lines indicate $T_{c1}$ and $T_{c2}$. In both cases, $\Gamma_3$ ($c$-axis magnetism) is associated with $T_{c1}$, and $\Gamma_1$ or $\Gamma_2$ (in-plane magnetism) with $T_{c2}$. Note that symmetry considerations described in the text indicate $\Gamma_2+\Gamma_3$ is the correct description of the low T state. Error bars indicate one standard deviation.}

\label{flo:Refinement}
\end{figure*}

The next step in determining the magnetic structure was fitting the neutron scattering intensity data to symmetry allowed structures with the given magnetic wave vector via Rietveld refinement. We used group-theoretical analysis to generate the irreducible representations ("irreps") of the little space group, which are shown in Table \ref{flo:IrrRepTable}. These irreps were computed by hand via the method outlined by Ref. \cite{Izyumov1979} (see Appendix \ref{app:GroupTheory} for these calculations), and were cross-checked with the program SARAh \cite{SARAh}. 
There were originally four basis vectors in the two-dimensional irrep $\Gamma_3$ treating the two magnetic ions in the unit cell separately. The basis vectors of $\Gamma_3$ were combined so as to preserve the equivalency of the two Ni sites. This site-equivalency is necessary to permit a second transition at $T_{c2}$ (see Appendix \ref{app:GroupTheory} for details).
We refined the elastic scattering data at $5\>$K (below the first transition) and at $2\>$K (below the second transition) using the Fullprof software package \cite{Fullprof} after subtracting the average of high temperature data acquired for temperatures between $12\>$K amd $20\>$K to isolate the temperature-dependent Bragg peaks. The space groups and their respective best fit $\chi^{2}$ values
are listed in Table \ref{flo:IrrRepTable}, and the refinements are shown in Fig. \ref{flo:Refinement}. In accord with the DFT results, we carried out the refinements assuming only one type of magnetic ion, and the resulting model fits the data quite well.




In refining the magnetic structure at $5\>$K, we used just one irrep at a time because the sample has only been cooled through one second-order phase transition at $5\>$K. $\Gamma_3$ yielded the best fit. For the $2\>$K data we fit to combinations of $\Gamma_3$ (the $5\>$K irrep) with $\Gamma_1$ and $\Gamma_2$ and found both combinations fit the $2\>$K data equally well (right two columns in Table \ref{flo:IrrRepTable}). To test this two-stage order, we repeated the refinements allowing multiple irreps at all temperatures. As Fig. \ref{flo:Refinement}(e)-(f) show, the relative weights of $\Gamma_1$ and $\Gamma_2$ refine to zero above $T_{c2}$, meaning that only $\Gamma_3$ is present for $T_{c1} < T < T_{c2}$.

The refined magnetic structure for temperatures between $T_{c1}$ and $T_{c2}$ [Fig. \ref{flo:Refinement}(d)] has all spins aligned along the $c$ axis, with the moment size modulated versus displacements within the basal plane and along the $c$ axis. This implies that some spins fluctuate more than others within this finite $T$ ordered phase. In the magnetic structure below $T_{c2}$ [Fig. \ref{flo:Refinement}(c)] every spin gains a counterrotating $ab$ plane component (where the two Ni spins in the unit cell rotate in opposite directions versus displacement) while the amplitude of the $c$-axis component continues to increase upon cooling. Thus we conclude that $T_{c1}=6.3(1)\>{\rm K}$ is associated with ordering the $c$-component of spins while the in-plane spin components only order for $T<T_{c2}=4.8(1)\>{\rm K}$.

Although neutron diffraction cannot distinguish in-plane spin structures based on $\Gamma_1$ and $\Gamma_2$, symmetry analysis identifies the one based on  $\Gamma_2$ as the correct low temperature  structure. This is because the addition of $\Gamma_1$ would not reduce the symmetry of the system, and therefore could not result in a phase transition at $T_{c2}$. Meanwhile, $\Gamma_2$ breaks a mirror-plane that is present in the $\Gamma_3$ structure so its appearance must be associated with a phase transition (see Appendix \ref{app:GroupTheory} for details). Therefore, we can identify $\Gamma_2$ as the proper in-plane magnetic structure. $\Gamma_2$ has ferromagnetic in-plane bond-dependent correlations (see Fig. \ref{flo:InPlaneStructure}). Although the magnetic structure breaks inversion symmetry, the counter-rotation precludes a definite handedness as seen in spiral incommensurate ferroelectrics \cite{Lawes2005}, so we do not expect ferroelectricity in this compound.

The peak widths in the refined model in Fig. \ref{flo:Refinement} were defined by the nuclear peak refinement (see Appendix \ref{app:CorrelationLength}), but the magnetic Bragg peaks are slightly wider than the peaks from the refined model. This indicates the magnetic correlation length is less than the correlation length of the nuclear structure. We can quantify this by fitting the 
0.81 \AA$^{-1}$ peak with a convolution of a Gaussian (with peak width defined by the nuclear phase) and a Lorentzian profile, 
where the inverse of the Lorentzian HWHM is the magnetic correlation length. Using this method, we infer a magnetic correlation length of  $152 \pm 16$ \AA. (See Appendix \ref{app:CorrelationLength} for details.) It is noteworthy that the spin correlations extend well beyond the correlation length anticipated for oxygen vacancies, consistent with uniform Ni$^{3+}$.

At $T=1.8\>$K, the refined ordered moments have a fixed in-plane magnitude while their $c$ axis component is spatially modulated [see Fig. \ref{flo:Refinement}(c)]. The overall size of the ordered moments range from 1.43 $\rm \mu_B$/Ni to 0.32 $\rm \mu_B$/Ni, with a mean value of $0.96 \> \rm \mu_B$/Ni. These values are taken from refinements which allow the magnetic peak width to be larger than the nuclear peak width so that all the elastic magnetic diffraction is accounted for. 
Adding this to the fluctuating moment from the inelastic sum-rule analysis above, we find that the total magnetic neutron scattering corresponds to a mean squared moment of $ m_{total}^2 =  m_{static}^2 + m_{dynamic}^2 = 4.2(7) \> \rm \mu_B^2/Ni$, which is slightly less than $ m_{\chi}^2 = 5.11(4) \> \mu_B^2/{\rm Ni}$ inferred from high-$T$ susceptibility data through Curie-Weiss analysis. We also find that $\frac{m_{dynamic}^2}{m^2_{total}} =$ 78(4)\% of the magnetism remains dynamic within the ordered phase ($T=2\>$K). 

Theoretically, the neutron spectral weight from elastic magnetic scattering is proportional to $\langle {\bf S} \rangle\langle {\bf S} \rangle$ and the total magnetic scattering is proportional to $ \langle {\bf S} \cdot {\bf S} \rangle$ \cite{Squires}, so that the ratio for dynamic vs total magnetic spectral weight for a fully static spin configuration is $\frac{\langle {\bf S}^2 \rangle - \langle {\bf S} \rangle^2 }{\langle {\bf S}^2 \rangle} = \frac{1}{S+1}$. 
So theoretically, with $S=3/2$ spins  $\frac{1}{3/2+1}  = \frac{2}{5}$ of the magnetic spectral weight should be dynamic in  $\rm NaNi_2BiO_{6-\delta}$. Our measured ratio $\frac{m_{dynamic}^2}{m^2_{total}} = 78(4)$\% is twice this, indicating the effects of a mixed $S=3/2$, $J=1/2$ state, possibly combined with frustration producing a more dynamic state than anticipated for a long range ordered or maximally frozen $S=3/2$ spin system. 

\section{Discussion \label{sec:Discussion}}

$\rm NaNi_2BiO_{6-\delta}$ has a larger magnetic unit cell and a more complex magnetic ground state than related Ni honeycomb compounds \cite{Seibel2013, Zvereva2015}. As we shall now show, the in-plane spin structure is particularly interesting, and points to bond dependent magnetic interactions.

Two-step longitudinal to transverse polarized long-wavelength magnetic ordering has been seen in other materials with easy axis anisotropy and competing interactions such as $\rm TbMnO_3$ \cite{Kenzelmann2005} and $\rm Ni_3V_2O_8$ \cite{Kenzelmann2006}, and Nagamiya provided a theoretical description of this phenomenon \cite{Nagamiya1967}. Association of the transitions in $\rm NaNi_2BiO_{6-\delta}$ with this mechanism  is supported by reports of a  Ni$^{2+}$ easy-axis anisotropy along $c$ in the honeycomb compounds ${\rm Na_3Ni_2SbO_6}$ and ${\rm Li_3Ni_2SbO_6}$ \cite{Zvereva2015}, which have similar Ni ligand environments to $\rm NaNi_2BiO_{6-\delta}$.
With an easy-axis anisotropy, one would expect low energy structure in the spin-wave spectrum at energy transfer of $T_{c1} - T_{c2}= 1.5\>$K, or $0.13\>$meV. However, our neutron experiment does not resolve the spectrum below $0.25\>$meV, so we could not detect such structure. 

\begin{figure}
\centering\includegraphics[scale=0.55]{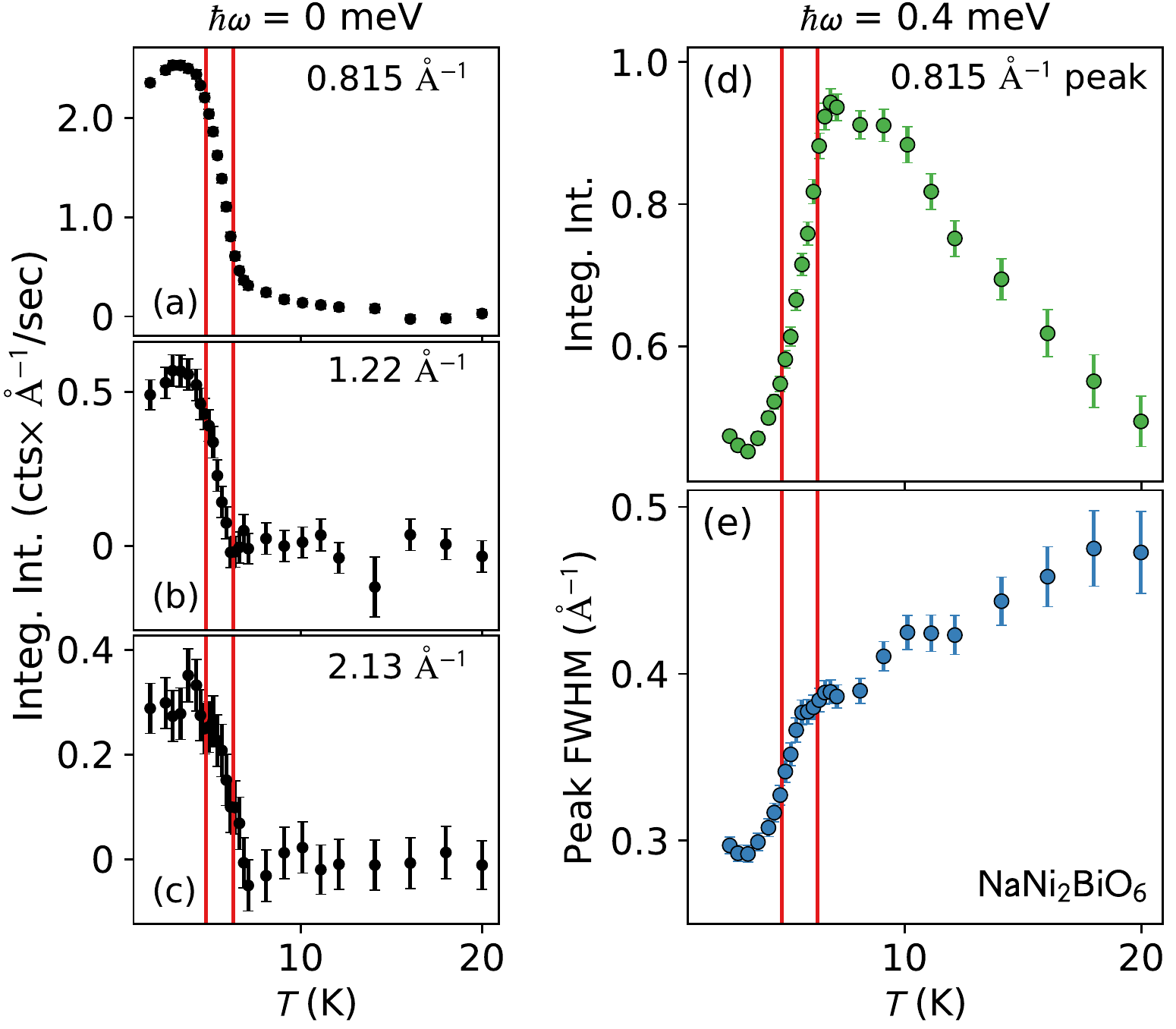}

\caption{Temperature dependence of magnetic peaks, extracted from Gaussian fits. (a)-(c) Order parameter curves for three elastic magnetic peaks. (d) and (e) respectively show the area and FWHM of the $\hbar \omega=0.5$ meV inelastic peak at 0.815 \AA$^{-1}$. (d) has the same units as (a-c). Red vertical lines indicate $T_{c2}$ and $T_{c2}$ from heat capacity. Error bars indicate one standard deviation.}

\label{flo:OrderParameter}
\end{figure}

One puzzling aspect of the magnetic order is the temperature-dependent elastic scattering [see Fig. \ref{flo:OrderParameter}(a)-\ref{flo:OrderParameter}(c)], which does not follow the typical single-exponent order-parameter curve for a second order transition. The 
magnetic Bragg diffraction intensity 
increases linearly as temperature decreases between $T_{c1}$ and $T_{c2}$, and then flattens off and decreases slightly at the lowest temperatures. This low-temperature decrease in elastic intensity is accompanied by an enhancement of inelastic fluctuations, revealed by the small upturn in Fig. \ref{flo:OrderParameter}(d).  This indicates a weakening of the counter-rotating spin order as might occur near a transition to a different phase. We leave this feature to be explored in future studies.

The observed ordering wave vector ${\bf q}=(\frac{1}{3},\> \frac{1}{3},\> 0.154 \pm 0.011)$ is unusual for honeycomb compounds; in fact unprecedented to our knowledge. The $(1/3, \>1/3)$ in-plane wave vector is difficult to stabilize on the honeycomb lattice, and suggests a highly frustrated set of exchange interactions. $(1/3, \>1/3)$ honeycomb order is found in phase diagrams of isotropic exchanges only in the "spiral phase" when $(J_1 - 2J_2)/(J_2-J_3) = 0.5$ exactly \cite{Li2012}. We consider this possibility unlikely because (i) it is stabilized in a vanishingly small region of parameter space, and (ii) the spiral phase in-plane structure is co-rotating, and does not match the counterrotating $\rm NaNi_2BiO_{6-\delta}$ in-plane structure. A better explanation for the $(1/3, \>1/3)$ structure, as we will explain shortly, is bond-dependent exchange interactions.

The long wavelength modulation along the $c$ axis requires competing interactions along $c$.
The exchange pathways for the first, second, and third nearest inter-plane neighbors are Ni-O-Na-O-Ni, which we expect to have $J \sim 0.1 \> {\rm meV}$ (by comparison to the same exchange pathway in $\rm NaNiO_2$ \cite{Meskine2005}). At the mean-field level, it is not possible to stabilize long-wavelength $c$-axis order with isotropic exchanges between only adjacent planes (see Appendix \ref{app:LuttingerTisza}).  
An inter-plane Dzialoszynski-Morya (DM) exchange ${\bf D} \cdot ({\bf S}_i \times {\bf S}_j)$ is allowed in this crystal  structure and would also tend to produce $c$-axis modulation (see Appendix \ref{app:Incommensurability} for details), but it only acts on in-plane moments and would not stabilize the intermediate temperature collinear magnetic structure.
The $c$ axis modulation requires a mechanism which stabilizes both ordered phases with the same wave-vector. Some possible mechanisms are (i) weak next nearest plane exchange competing with the nearest-plane exchange  \cite{Kenzelmann2006}, (ii) an interplane biquadratic exchange $J({\bf S}_i \cdot {\bf S}_j)^2$ competing with a ferromagnetic Heisenberg interplane exchange, or (iii) exchange disorder from oxygen deficiencies in some cases might be able to favor a modulated state \cite{Scaramucci_2018}. Any of these could produce the observed long-wavelength modulation along $c$.

 
We also note that in the quasi-2D hydrate version of $\rm NaNi_2BiO_{6-\delta}$ ($\rm NaNi_2BiO_{6-\delta} \cdot 1.7 H_2O$  with $\rm H_2O$ molecules in-between the planes) the transition temperature as determined by  heat capacity ($T_{c}=5\>$K) is isimilar to the 5.6 K average of $T_{c1}$ and $T_{c2}$ for anhydrate though there is a single broad transition for the hydrate \cite{Seibel2014}. This suggests that inter-plane interactions are not very significant and the ordered magnetism is quasi-2D even for the anhydrate. According to the Mermin-Wagner theorem, magnetic order in a 2D system requires anisotropic interactions. 



\begin{figure}
\centering\includegraphics[scale=0.4]{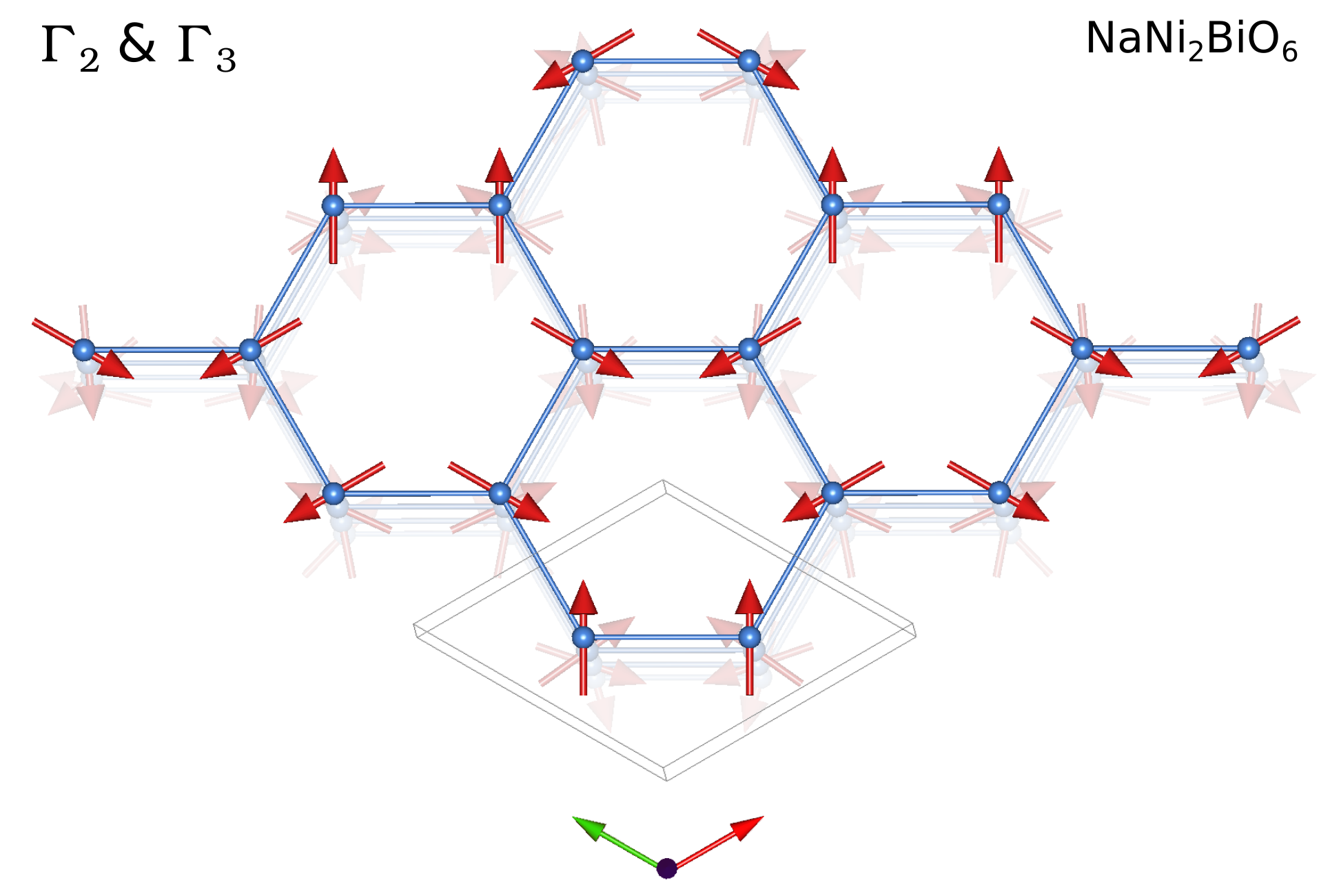}

\caption{In-plane magnetic structure of $\rm NaNi_2BiO_{6-\delta}$ below $4\>$K  as described by IR $\Gamma_2$. The lightly-shaded structures in the background show how the spins in subsequent planes within one chemical unit cell are aligned.}

\label{flo:InPlaneStructure}
\end{figure}

Perhaps the most intriguing aspect of the magnetic order is the counterrotating in-plane structure, shown in Fig. \ref{flo:InPlaneStructure}. The out-of-plane magnetic correlations are clearly antiferromagnetic, indicating an 
antiferromagnetic nearest neighbor exchange---but the in-plane correlations evidence a subtle sub-dominant interaction at play.
This in-plane structure is unusual because the the mean field component of isotropic exchange interactions average to zero for such structures. Specifically,  $\sum_{\langle ij \rangle} \langle{\bf S}_i\rangle \cdot \langle{\bf S}_j\rangle=0$ for nearest neighbor, next-nearest neighbor, and all further neighbor spin pairs forming a (1/3, 1/3) counterrotating spin state on the honeycomb lattice. This can be proved as follows: 
Fig. \ref{flo:InPlaneStructure} shows the angles between nearest-neighbor spins are always 0\textdegree\, 120\textdegree\, and 240\textdegree . Thus, for nearest neighbor exchange on any site,  $\sum_{\langle ij \rangle} \langle{\bf S}_i\rangle \cdot \langle{\bf S}_j\rangle = \cos 0^{\circ} + \cos 120^{\circ} + \cos 240^{\circ} = 0$. Extending this analysis to further neighbors is straightforward and yields the same result. (This result holds for other layers where the spins are rotated about the $c$ axis as shown in the lightly-shaded structures in Fig. \ref{flo:InPlaneStructure}.) 
This means the magnetic structure that we provide evidence for cannot be stabilized by isotropic exchange interactions at the mean-field level. 
This condition holds for each bond even if the three-fold axis is broken and the three bond directions have different interaction strengths, as in ${\rm Na_3Ni_2SbO_6}$ \cite{Zvereva2015}, because $\sum_{\langle ij \rangle} \langle{\bf S}_i\rangle \cdot \langle{\bf S}_j\rangle =0$ for each of the three distinct bond directions considered as groups. Confirming this conclusion is the fact that this structure is not found in theoretical phase diagrams for isotropic exchange interactions on the ideal honeycomb lattice \cite{Albuquerque2011, Li2012, Clark2011}.
In-plane Dzyaloshinskii-Moriya (DM) interactions are forbidden on the Honeycomb lattice because the midpoint between magnetic ions is a point of inversion \cite{Moriya}. This leaves two possibilities: either oxygen vacancy disorder influences the magnetic interactions in such a way as to stabilize this structure (DM interactions are allowed on bonds with oxygen vacancies), or there must be more exotic anisotropic interactions at play.




Certain anisotropic exchange interactions are possible through bond-dependent orbital interactions. In $\rm NaNi_2BiO_{6-\delta}$, the $(1/3, \>1/3)$ in-plane magnetic structure is consistent with two theoretical models: (i) a Kitaev-$\Gamma$-Heisenberg ($\rm K \Gamma H$) exchange with a negative Kitaev and off-diagonal $\Gamma$ terms producing a 120\textdegree\ ordered structure \cite{Rau2014,Chaloupka2015}, and (ii) a different bond-dependent exchange called a 120\textdegree\ compass model exchange \cite{CompassModelReview, Mostovoy2002} (this interaction is analogous to the Kitaev bond-dependent interaction, but the Ising-like exchange directions are coplanar and 120\textdegree\ apart). Both these models produce the observed in-plane structure on the honeycomb lattice \cite{Wu2008, Zhao2008, Nasu2008}, and either case implies strong bond-dependent exchange in $\rm NaNi_2BiO_{6-\delta}$.

A similar modulated counterrotating magnetic order was observed in honeycomb $\rm \alpha \text{-}Li_2Ir0_3$ with ordering wave vector ${\bf q}=(0.315(9), 0, 0)$ \cite{Williams2016}, or ${\bf q}=(0.156(5), 0.156(5), 0) \approx (1/6, 1/6, 0)$ expressed in the $\rm NaNi_2BiO_{6-\delta}$ reciprocal lattice. In this case, the spin structure is attributed to a Kitaev-like Hamiltonian with different couplings on the vertical and zig-zag bond directions \cite{Biffin2014, Williams2016, Itmar2016}. Although $\rm \alpha \text{-} Li_2IrO_3$ and $\rm NaNi_2BiO_{6-\delta}$ share a 1/3 counterrotating structure, there are important differences. First, $\rm NaNi_2BiO_{6-\delta}$ has two magnetic phase transitions and $\rm \alpha \text{-} Li_2IrO_3$ has one. Second, the counterrotating structures are different and the $\rm \alpha \text{-} Li_2Ir0_3$ spin structure is inconsistent with theoretical predictions from the $\rm K \Gamma H$ or 120\textdegree\ compass model. Third, $\rm \alpha \text{-} Li_2Ir0_3$ does not have three-fold rotation symmetry about its magnetic sites, and its structure requires either $K_z \neq K_x, K_y$ or an additional Ising term on the $K_z$ bonds to stabilize the counterrotating order \cite{Williams2016}. Meanwhile, $\rm NaNi_2BiO_{6-\delta}$ can be explained by a Hamiltonian that preserves the three-fold axis.

\begin{figure}
\centering\includegraphics[scale=0.15]{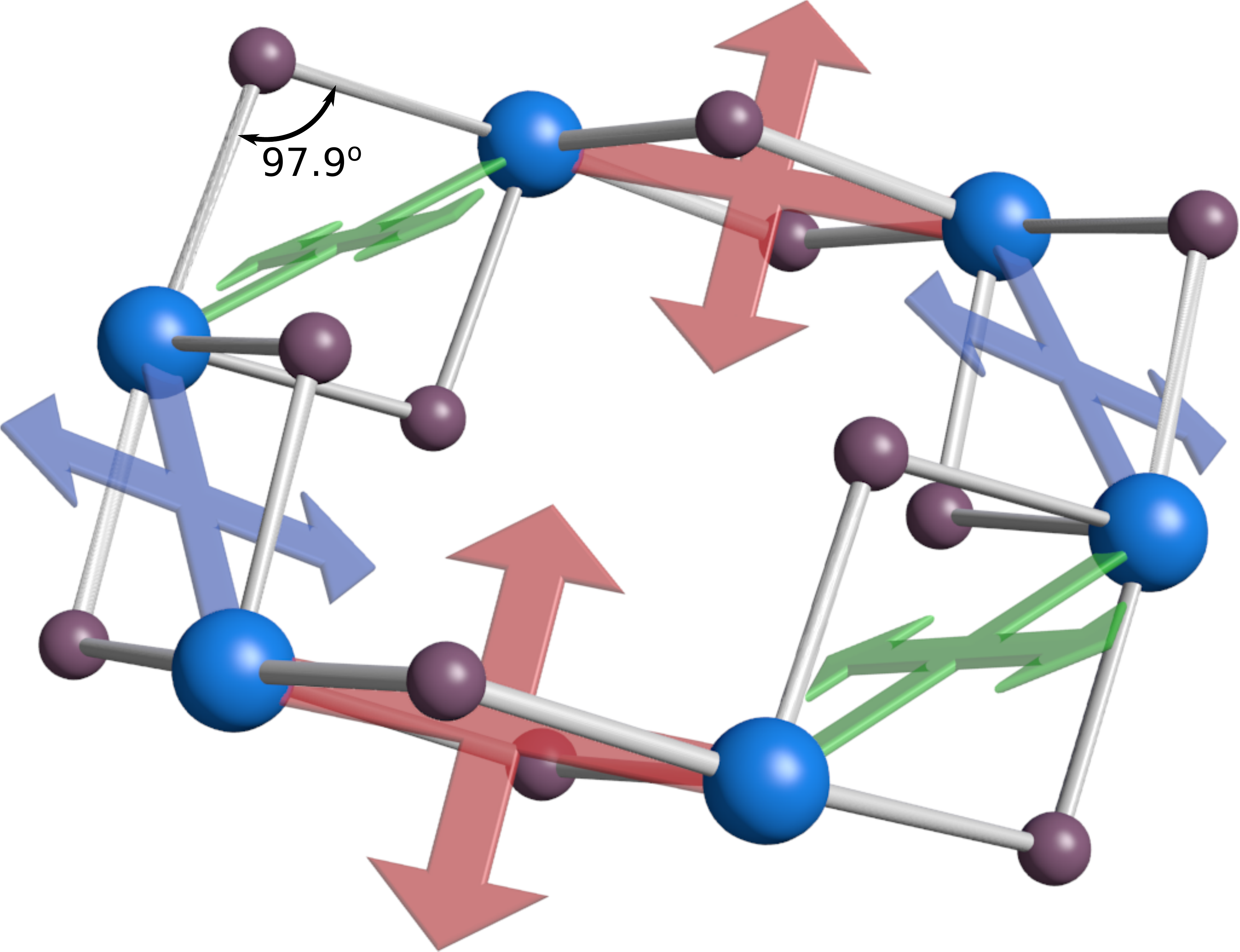}

\caption{Ni-O-Ni exchange pathways in $\rm NaNi_2BiO_{6-\delta}$. The Ni-O-Ni bond angle is 97.9(4)\textdegree\, leading to anisotropic exchange perpendicular to the ligand-ion plane (shown by the red, green, and blue arrows) plus an isotropic term.}

\label{flo:ExchangePathways}
\end{figure}

The $c$-axis component of the magnetic wave vector for  $\rm NaNi_2BiO_{6-\delta}$ is indistinguishable from $\ell=1/6$. Curiously, the  $(\frac{1}{3},\> \frac{1}{3})$ 120\textdegree\ ordered structure in Fig. \ref{flo:InPlaneStructure} has a six-fold degeneracy in its ground state: the spins can be rotated 60\textdegree\ (opposite directions for the two Ni sites) and the structure is related by a global translation and rotation of the axes (i.e., energetically equivalent). This six-fold degeneracy will give rise to a local minimum when $\ell=1/6$ such that the system explores all the degenerate in-plane states, which may play a part in stabilizing the magnetic order with a wave-vector close to $\ell=1/6$. 


The microscopic origin of the bond-dependent exchange in $\rm NaNi_2BiO_{6-\delta}$ cannot be determined from the data and calculations reported here, but we present some possibilities: 
In the case of large spin orbit coupling and a 90\textdegree\ ion-ligand-ion bond, the nearest neighbor exchange is an Ising-like anisotropic exchange oriented perpendicular to the plane formed by the superexchange pathway \cite{Jackeli2009}. This effect emerges also for ions with intermediate spin-orbit coupling, such as Ru$^{3+}$ in $\rm RuCl_3$ \cite{Kim2015, Banerjee2016}, due to direct overlap of $d$ orbitals \cite{Kim2015}. 
Such bond-dependent effects are not limited to Ir and Ru; they have been predicted also for high-spin $d^7$ ions \cite{Cobalt_Liu2018,Cobalt_Sano2018}, which are electronically equivalent to high-spin Ni$^{3+}$. 
In addition, the orbital mixing of the O ligands with the heavy Bi ion produces the effect of strong Ni spin orbit coupling \cite{stavropoulos2019microscopic}, enhancing the Ni bond-dependent interactions.

In $\rm NaNi_2BiO_{6-\delta}$, the situation is imperfect with a 97.9(4)\textdegree\  Ni-O-Ni bond (shown in Fig. \ref{flo:ExchangePathways}), 
so that other exchange terms are present: the nearest neighbor exchange can be written $J_{\parallel}{\bf S}_{1\>\parallel}{\bf S}_{2\>\parallel} + J_{\perp}{\bf S}_{1\>\perp}{\bf S}_{2\>\perp}$ ($\parallel$ and $\perp$ denote the directions in and perpendicular to the Ni-O-Ni plane) and $J_{\perp}>J_{\parallel}$. 
The resulting exchange anisotropies, shown in  Fig. \ref{flo:ExchangePathways}, are rotated 38.5\textdegree\ out of the plane so that the anisotropy directions are 94.3(5)\textdegree\ apart ($\theta'=94.3(5)^{\circ}$ in the nomenclature of ref. \cite{Zou2016}), making this exchange very close to the celebrated Kitaev model where $\theta'=90^{\circ}$. According to recent theoretical work \cite{Zou2016}, $87^{\circ}<\theta'<94^{\circ}$ is the range of Kitaev spin liquid behavior, so that $\rm NaNi_2BiO_{6-\delta}$  may be  right on the boundary between Kitaev and 120\textdegree\ compass behavior. (However, this boundary is almost certainly shifted in the presence of non-Kitaev exchange as in this material.) 
The in-plane structure can be explained on either side of the $\theta'$ critical point, but given the presence of Heisenberg and of off-diagonal exchange, it may be more appropriate to associate this material with the $\rm K \Gamma H$ model. 

Interestingly, both components of the mixed spin-orbital state support Kitaev interactions and 120\textdegree\ order.
For $S=3/2$, it is a bond-dependent Kitaev interaction with a tiny $\Gamma$  due to three holes in $d$-orbitals, i.e., $d^7$ \cite{stavropoulos2019microscopic}.
For $J=1/2$, it is again a bond-dependent Kitaev interaction with a small $\Gamma$  \cite{Cobalt_Liu2018}.
Either way, Kitaev is a dominant interaction, so we fully expect the mixed $S=3/2$, $J=1/2$ to have dominant Kitaev exchange.

If this is true, we can expect to find the exotic quasiparticles of the Kitaev model in $\rm NaNi_2BiO_{6-\delta}$. It has been shown theoretically that the Kitaev model for half-integer spin (including $S=3/2$ and $J=1/2$, the components of the mixed state) has Majorana fermion excitations \cite{Baskaran2008}, and it is believed that the Kitaev entropy plateau is associated with a plaquette valence-bond state \cite{Oitmaa2018}. This suggests  that the region above the ordering transition where the $Q-E$ dependence of magnetic neutron scattering is distinct from that in the ordered state [Fig. \ref{flo:Neutronscans}(c)] could be associated with emergent Majorana physics. 

\section{Conclusion}

We have acquired and analyzed heat capacity, ESR, and neutron scattering data on $\rm NaNi_2BiO_{6-\delta}$, and all evidence points toward Kitaev-like bond-dependent exchange in this compound. Heat capacity shows missing entropy consistent with an incipient entropy plateau of a high-Spin Kitaev model.  ESR data and DFT calculations indicate Ni$^{3+}$ is in a mixed $S=3/2$, $J=1/2$ state with a thermally populated low spin $S=1/2$ state. All this comports with theoretical predictions for $d^7$ Kitaev exchange in the honeycomb geometry. Neutron scattering indicates a two-stage magnetic order with substantial short ranged magnetic correlations in the paramagnetic phase, and inelastic scattering shows strong quantum fluctuations within the ordered phase. 
The observed magnetic structure has unusual counterrotating in-plane correlations, which are not favored by isotropic interactions but are favored by bond-dependent exchange. The special ligand environment and in-plane structure inferred from diffraction data is consistent with the 120\textdegree\ phase of the $\rm K \Gamma H$ model.

These results are significant firstly because bond-dependent interactions  in Ni have not previously been documented; conventional wisdom says its weaker spin-orbit coupling would render bond-dependent effects too weak to impact magnetism \cite{Jackeli2009}. But in $\rm NaNi_2BiO_{6-\delta}$, the effect is significant possibly as a consequence of covalent bonding of superexchange mediating oxygen orbitals with Bi orbitals that are subject to strong spin-orbit coupling. This raises the possibility of discovering Kitaev-like spin-liquid phases in $3d$ transition metal oxides with edge sharing six-fold coordination. Secondly, the observation of Kitaev physics in a mixed $S=3/2$, $J=1/2$ compound raises the possibility in such materials of observing new kinds of quasiparticles which have been predicted for high-spin Kitaev models \cite{Baskaran2008, Oitmaa2018, stavropoulos2019microscopic}.

\section*{Acknowledgments}

This work was supported as part of the Institute for Quantum Matter, an Energy Frontier Research Center funded by the U.S. Department of Energy, Office of Science, Basic Energy Sciences under Award No. DE-SC0019331.  This work was also partly supported by the Natural Sciences and Engineering Research Council of Canada and the Center for Quantum Materials at the University of Toronto. Computations were performed on the Niagara supercomputer at the SciNet HPC Consortium. SciNet is funded by: the Canada Foundation for Innovation under the auspices of Compute Canada; the Government of Ontario; Ontario Research Fund - Research Excellence; and the University of Toronto. AS and CB were supported through the Gordon and Betty Moore foundation under the EPIQS program GBMF4532. Access to MACS was provided by the Center for High Resolution Neutron Scattering, a partnership between the National Institute of Standards and Technology and the National Science Foundation under Agreement No. DMR-1508249. We also acknowledge helpful discussions with Kemp Plumb and Oleg Tchernyshyov.

%


\newpage 

\quad

\pagebreak

\appendix
 
\renewcommand{\thefigure}{A\arabic{figure}}
\renewcommand{\thetable}{A\arabic{table}}
\setcounter{figure}{0}


\section{Heat Capacity}

\subsection{Magnetic Entropy from Phonon Subtraction}\label{app:PhononSubtraction}

\begin{figure}[H]
\centering\includegraphics[scale=0.44]{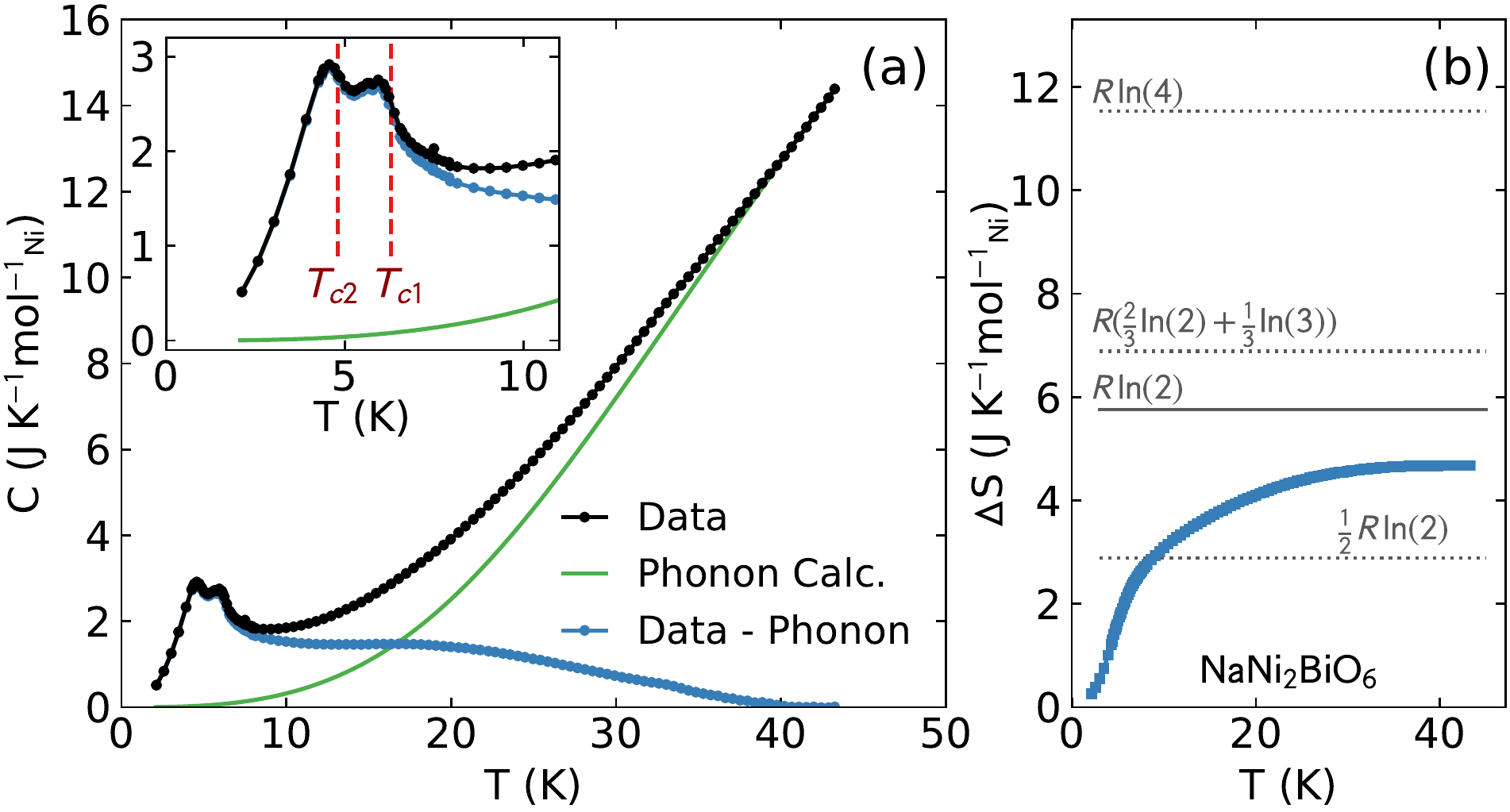}

\caption{Low temperature heat capacity of $\rm NaNi_2BiO_{6-\delta}$. (a) Plot of measured heat capacity (black) and calculated phonon background (green). The blue data show the measured heat capacity minus the calculated background $\Delta C$, which is an estimate of the magnetic contribution to heat capacity. The inset shows two transitions at $T_{c1}=6.3\>{\rm K}$ and $T_{c2}=4.8\>{\rm K}$. (b) Entropy obtained from integrating heat capacity (extrapolated to zero using a $T^3$ fit).}

\label{flo:HeatCapacity_b}
\end{figure}

No nonmagnetic analogue to  $\rm NaNi_2BiO_{6-\delta}$ is currently available to measure the phonon specific heat and isolate the magnetic contribution to heat capacity in $\rm NaNi_2BiO_{6-\delta}$. Therefore, we attempted to estimate the magnetic entropy by subtracting a phonon background calculated using the Debye equation for heat capacity
$c_v = 9nk_B \Big(\frac{T}{\Theta_D}\Big)^3 \int^{\Theta_D/T}_{0} \frac{x^4 e^x dx}{(e^x-1)^2}$ \cite{AshcroftAndMermin}. 
Here $n$ and $\Theta_D$ were fitted using the ten highest temperature data points (under the assumption that specific heat is lattice only by 40~K), which gave values of $n=2.90(6)$ per unit cell and $\Theta_D=206(2)\>$K. The results are shown in Fig. \ref{flo:HeatCapacity_b}, and indicate that between $2\>$K and $40\>$K the entropy only reaches 65\% of the originally proposed $\Delta S=R (2/3\ln(2) +1/3 \ln(3))$ [see Fig. \ref{flo:HeatCapacity_b}(b)].

In the Dulong-Petit limit $n$ should be 5 (the number of atoms per Ni). Our fitted value is 2.90(6). This discrepancy is a sign that the Debye estimate for heat capacity is unrealistic. Therefore, we do not have much confidence in the entropy computed from this background subtraction, and leave the presence of high temperature magnetic entropy as an open question.

\subsection{Identifying phase transitions}\label{app:PhaseTransitionIdentification}

\begin{figure}[H]
\centering\includegraphics[width=0.47\textwidth]{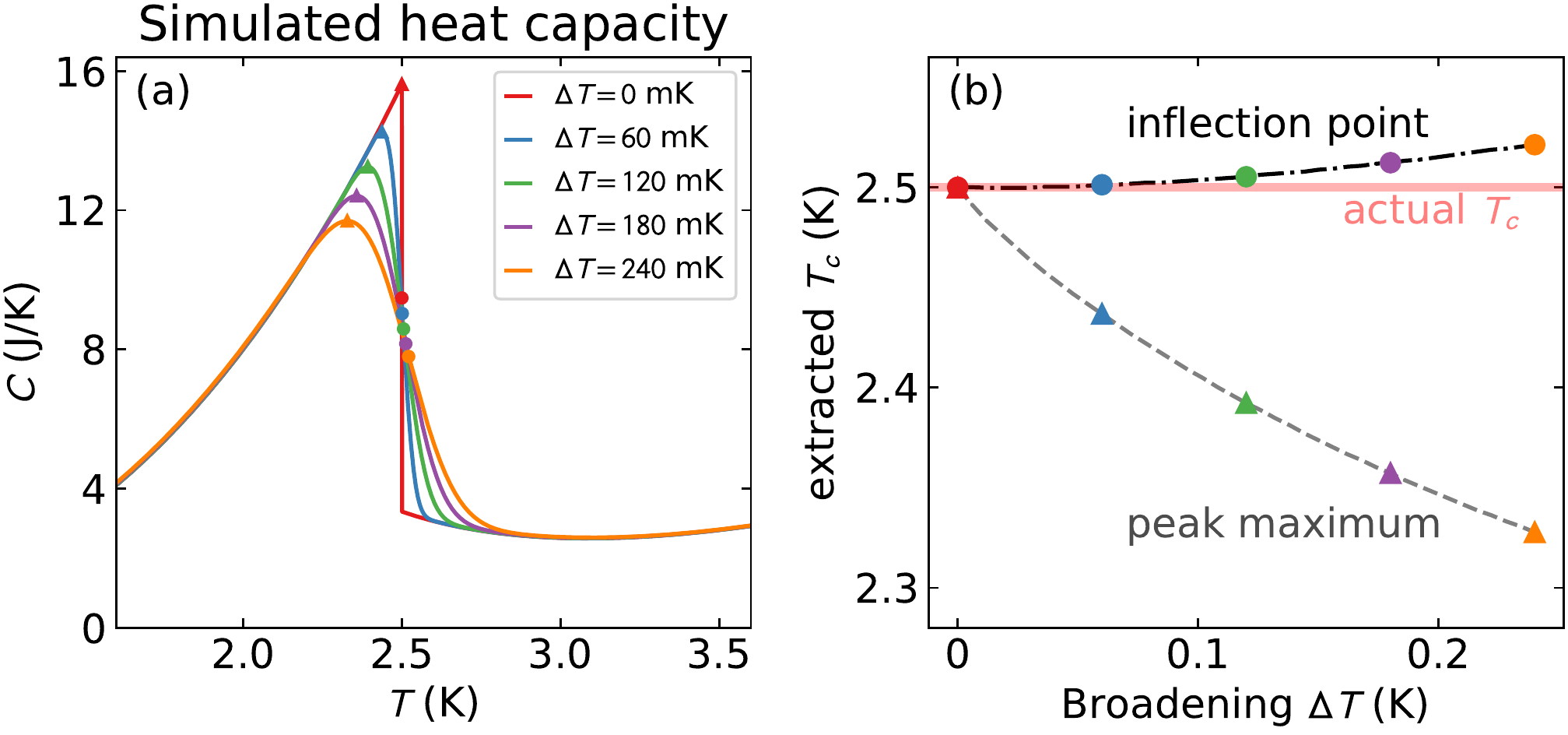}

\caption{(a) Simulated lambda anomaly in heat capacity convoluted with Gaussian profiles of varying widths to simulate experimental broadening. (b) Transition temperature $T_c$ extracted from the peak maximum (triangles) and the inflection point on the high-temperature side (circles). At all levels of broadening, the inflection point is a much better indicator of $T_c$.}

\label{flo:HeatCapacityPeak}
\end{figure}

The transition temperatures $T_{c1}$ and $T_{c2}$ are identified with the inflection point in the heat capacity peaks (where $\frac{d^2C}{dt^2}$ changes sign). Theoretically, a second order phase transition has a lambda discontinuity in the value of heat capacity, where the transition temperature is right at the discontinuity. Experimentally, these lambda anomaly peaks get smeared out in temperature because heat capacity is measured over a finite temperature range, because thermal equilibrium is not perfect, and because of slight sample inhomogeneities, etc. If one imagines a perfect lambda anomaly broadened in temperature, the transition temperature is no longer at the discontinuity peak (because the discontinuity is broadened), but can be reliably identified by the inflection point where the second derivative with respect to temperature changes sign (see Fig. \ref{flo:HeatCapacityPeak}). This is what we have identified as the critical temperature in our heat capacity data. 

\section{Density Functional Theory}\label{app:DFT}

We determined the valence of Ni by computing the partial density of states (PDOS) of $\rm NaNi_2BiO_6$, and $\rm NaNi_2BiO_{5.66}$ using the OPENMX ab-initio package. OPENMX \cite{ozakiopen} is a density functional theory code based on the linear combination of psudo-atomic orbitals formalisim \cite{Ozaki2003}. The exchange-correlation potential used is the Perdew-Burke Ernzerhof generalized gradient approximation \cite{Perdew1996}. An energy cutoff of 400 Ry is used for real-space integrations and a $8\times 8 \times 8$ $k$ grid samples the Brillouin zone. ($k$ is momentum with $\hbar = 1$.)

\begin{figure}
\centering\includegraphics[trim={1.3cm 2cm 0 0},clip,scale=0.31]{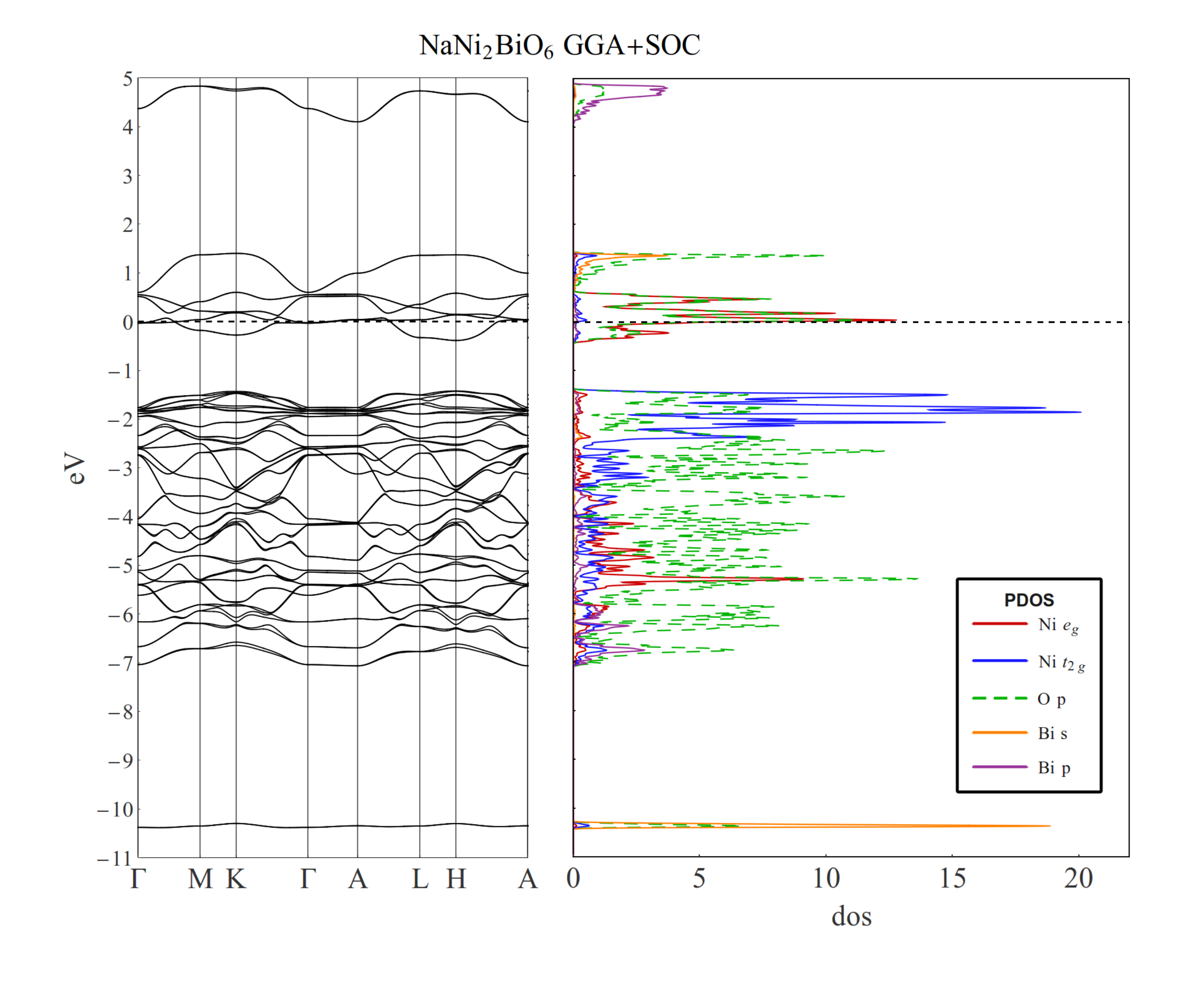}
\caption{PDOS of $\rm NaNi_2BiO_{6}$. The left panel shows the band structure and the right panel shows the density of states for the various orbitals.}
\label{flo:PDOS1}
\end{figure}

\begin{figure}
\centering\includegraphics[trim={1.3cm 2cm 0 0},clip,scale=0.31]{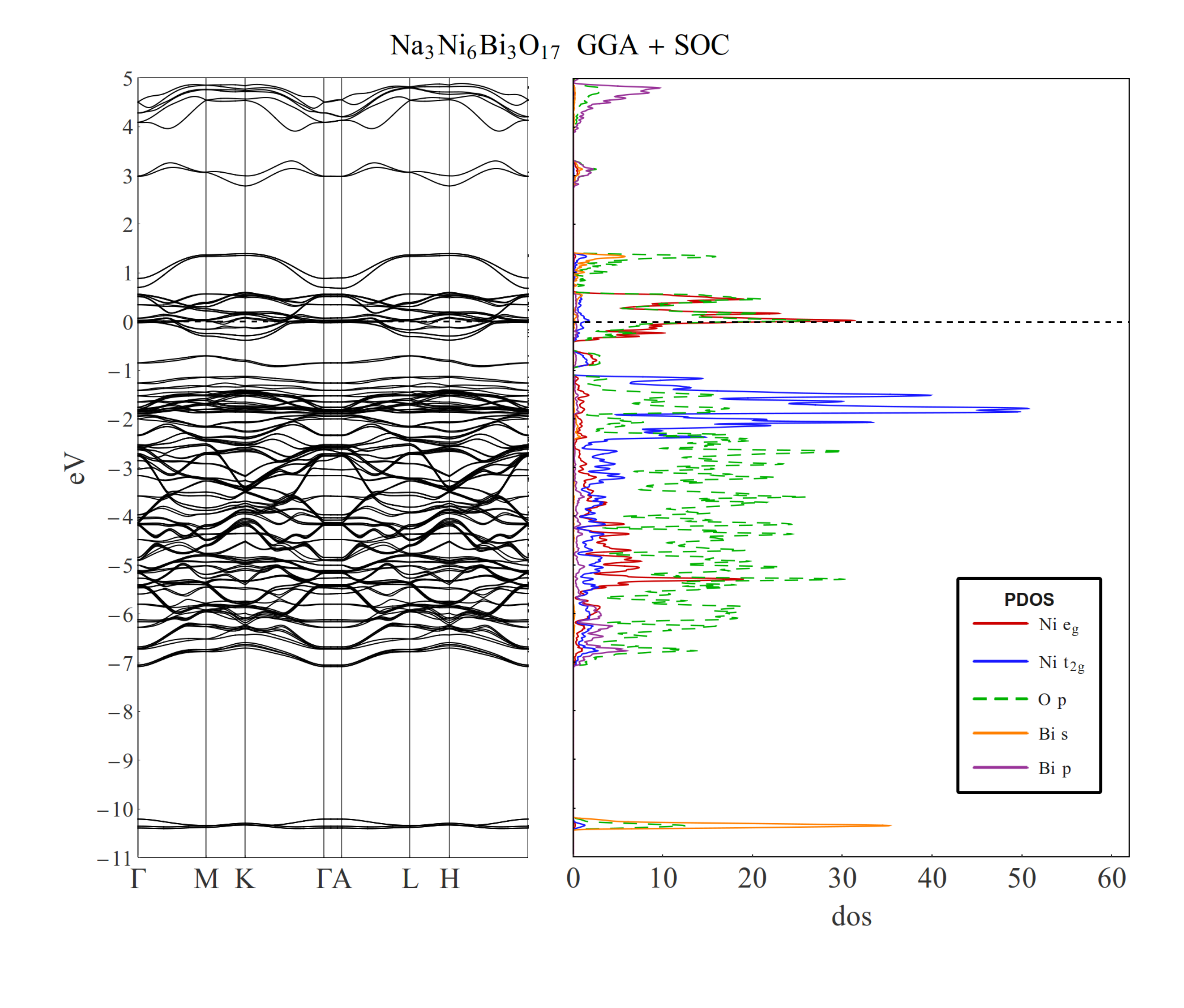}
\caption{PDOS of $\rm Na_3Ni_6Bi_3O_{17}$, which simulates $\rm NaNi_2BiO_{5.66}$. The left panel shows the band structure and the right panel shows the density of states.}
\label{flo:PDOS2}
\end{figure}

For $\rm NaNi_2BiO_6$ (Fig. \ref{flo:PDOS1}), there is one band with mainly Bi $s$-orbital character deep below the Fermi energy  around -10.5~eV, while one  O $p$-orbital band appears above the Ni $e_g$-orbitals, leading to 1/4-filling of the Ni $e_g$-orbitals ($d^7$).
To understand the valence of Ni in $\rm NaNi_2BiO_{5.66}$, we first triple the unit cell of $\rm NaNi_2BiO_{5.66}$ and remove one oxygen  to simulate  $\rm Na_3Ni_6Bi3O_{17}$. The PDOS of $\rm Na_3Ni_6Bi_3O_{17}$ (Fig. \ref{flo:PDOS2}) shows three Bi $s$-orbitals near -10.5~eV, but only two O $p$-orbitals above the $e_g$-bands, and one $p$-orbital below the $e_g$-bands. This charge redistribution maintains quarter filled $e_g$-orbitals of Ni and leads to the same $d^7$ configuration as $\rm NaNi_2BiO_{6}$. Note that the $e_g$-bands are heavily mixed with O $p$-orbitals near the Fermi energy, suggesting that indirect hopping paths are important in determining a microscopic spin model.

\begin{figure}
\centering\includegraphics[trim={1.3cm 2cm 0 0},clip,scale=0.31]{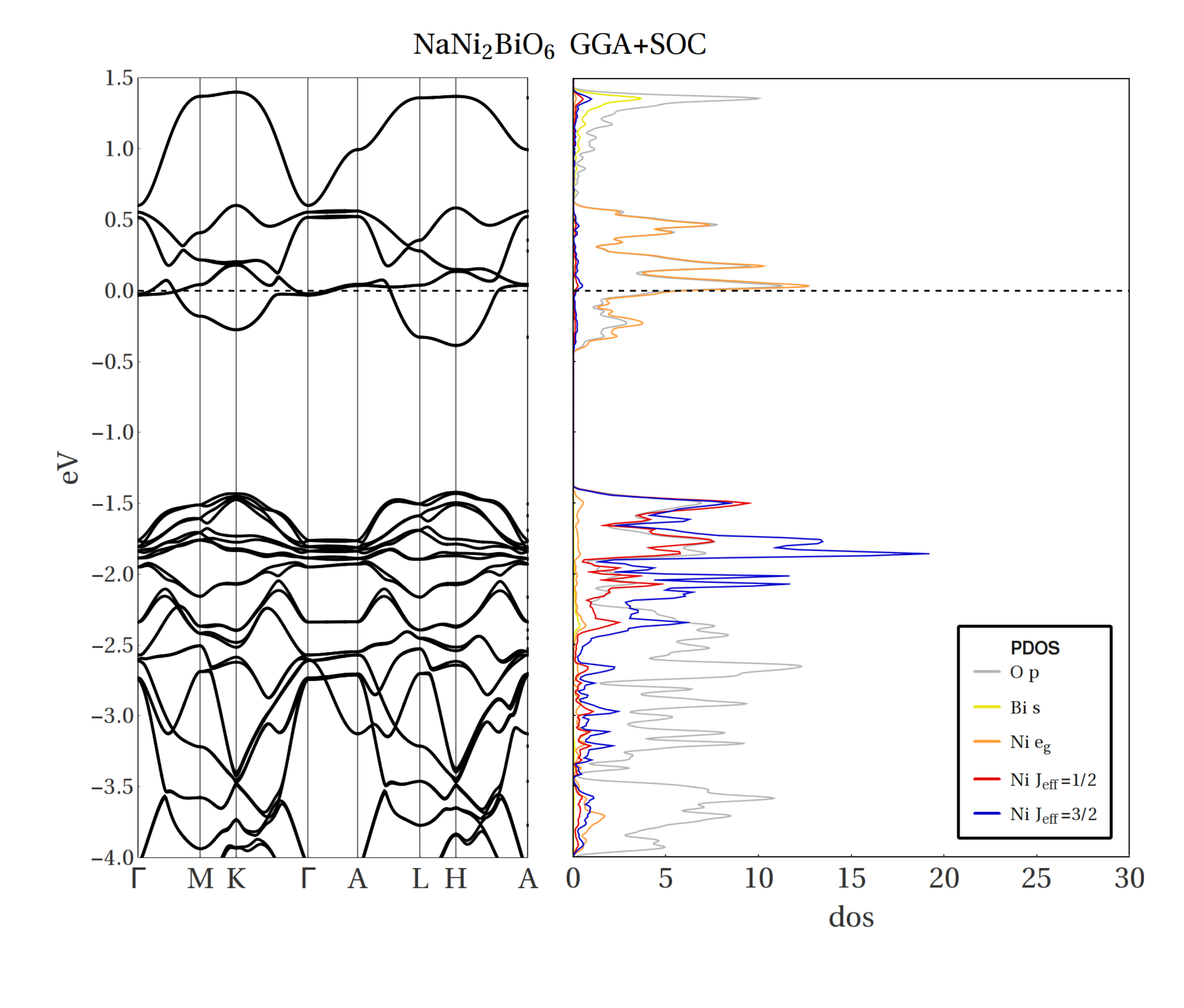}
\caption{PDOS of $\rm NaNi_2BiO_{6}$ projected into the $J=1/2$ and $S=3/2$ basis, showing that they are not well separated. The left panel shows the band structure and the right panel shows the density of states.}
\label{flo:PDOS3}
\end{figure}

To investigate the single-ion orbital state of Ni$^{3+}$, we projected the density of states to the $J=1/2$ and $S=3/2$ basis, shown in Fig. \ref{flo:PDOS3}. We find that the $J=1/2$ and $S=3/2$ are not well separated, and thus $\rm NaNi_2BiO_{6-\delta}$ appears to be in an intermediate state between $J=1/2$ and $S=3/2$.
Treating the trigonal distortion is smaller than SOC, a microscopic spin model of $\rm KJ \Gamma$ model for $J=1/2$ is dervied in ref. \cite{Jackeli2009}, and the observed spiral order is found in the regime of ferromagnetic Kitaev and small $\Gamma$ term, with a small antiferromagnetic $\Gamma'$ term induced by the trigonal distortion.

\section{Local moment computation}\label{app:localmoment}

To understand the size of local moment of $d^7$ electrons in $3d$ systems with comparable strengths of SOC and trigonal crystal field
splitting (CFS), we consider a site of $d$-orbitals surrounded by an octahedral environment.
The on-site Hamiltonian is modelled by the Kanamori interaction\cite{kanamori1963}, the CFS and the SOC:
\begin{eqnarray}
H & =  &  U \sum\limits_{\alpha} n_{\alpha\uparrow} n_{\alpha\downarrow} \nonumber
+ \dfrac{U'}{2} \sum\limits_{\substack{\alpha\neq\beta,\\\sigma,\sigma'}} n_{\alpha\sigma} n_{\beta\sigma'}  \\ \nonumber
& & - \dfrac{J_{H}}{2}\sum\limits_{\substack{\alpha\neq\beta,\\\sigma,\sigma'}} c^{\dagger}_{\alpha\sigma}  c^{\dagger}_{\beta\sigma'} c_{\beta\sigma} c_{\alpha\sigma'} 
+ J_{H}\sum\limits_{\alpha\neq\beta} c^{\dagger}_{\alpha\uparrow} c^{\dagger}_{\alpha\downarrow} c_{\beta\downarrow} c_{\beta\uparrow} \\ \nonumber
& & + \Delta_{oct}\sum\limits_{\substack{\alpha\in e_g,\\\sigma}}n_{\alpha\sigma}  + \Delta_{trig}\sum\limits_{\substack{\alpha,\beta\in t_{2g},\\\sigma}}c^{\dagger}_{\alpha\sigma} (D^{trig})_{\alpha\beta}\ c_{\beta\sigma} \\ 
& &
+ \lambda \mathbf{l} \cdot \mathbf{s}\ ,\ \
D^{trig}=\begin{bmatrix}
 0 & 1 & 1 \\
 1 & 0 & 1 \\
 1 & 1 & 0 
\end{bmatrix}
\label{eq:onsitHAM}
\end{eqnarray}
where the density operator $n_{\alpha\sigma}$ is given by $ c^{\dagger}_{\alpha \sigma} c_{\alpha \sigma}$, and
$c^{\dagger}_{\alpha\sigma}$ is the creation operator with $\alpha$ orbital and spin $\sigma$. 
$U$ and $U'$ are the intra-orbital and inter-orbital density-density interaction respectively, and $J_{H}$ is the Hund's coupling for the spin-exchange and pair-hopping terms. $\Delta_{\mathrm{oct}}>0$ is the octahedral CFS strength separating $e_{g}$ and $t_{2g}$ orbitals and $\Delta_{\mathrm{trig}}$ is the subleading CFS due to trigonal distortion, where $\Delta_{\mathrm{trig}}>0$ describes compressive 
distortion.
Operators $\mathbf{l}$ and $\mathbf{s}$ respectively denote angular momentum and spin for orbital $\alpha$ and spin $\sigma$, and $\lambda$ denotes the strength of SOC.

\begin{figure}

\centering\includegraphics[width=0.98\linewidth]{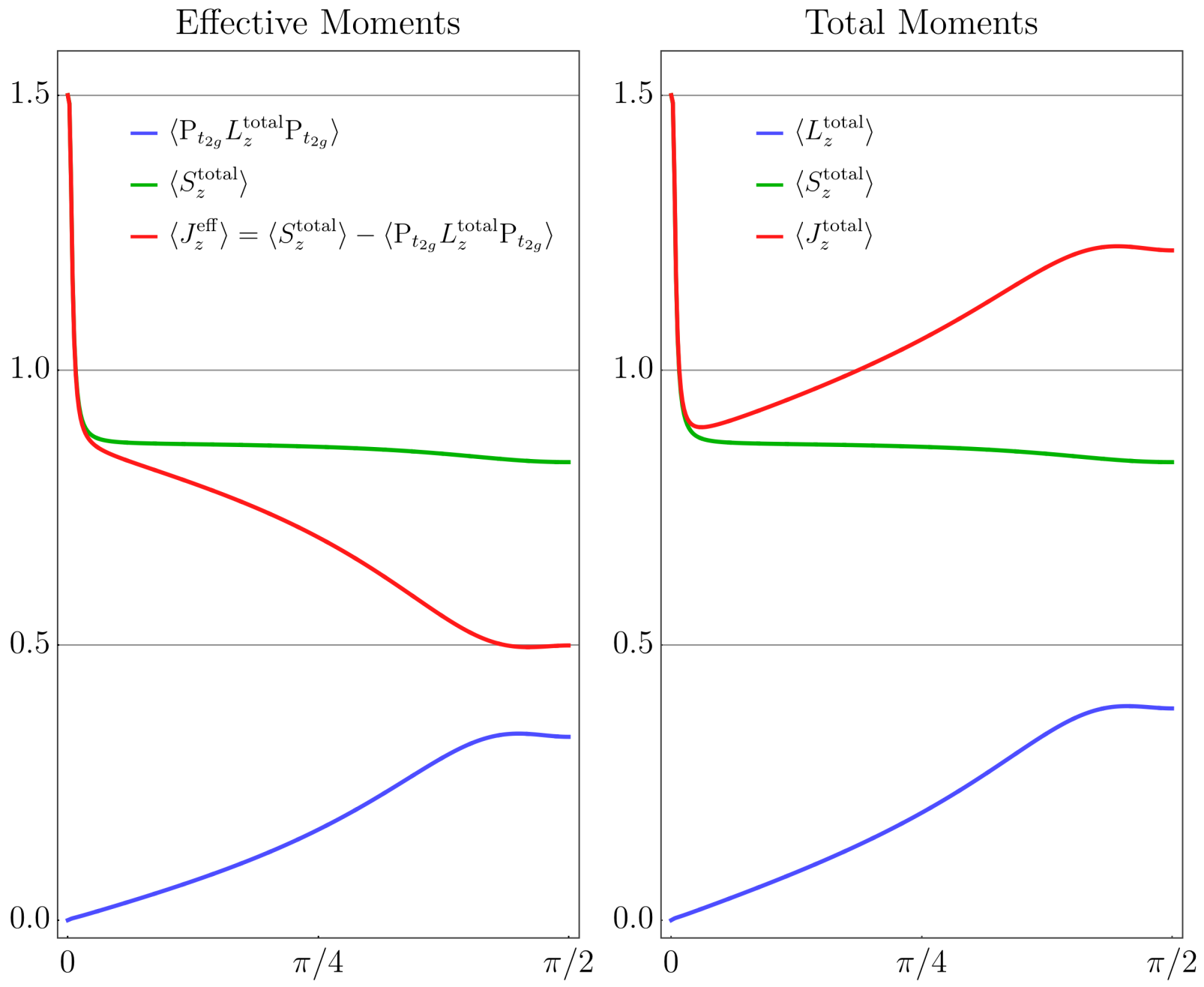}
  \caption{Local moment calculation for Ni$^{3+}$ $d^7$ as a function of $\phi$. The left panel shows the effective moment projecting out the $e_g$ orbitals, and the right panel shows the total moment including $e_g$ orbitals.}
    \label{fig:moments}
\end{figure}

We diagonalize the Hamiltonian (\ref{eq:onsitHAM}) for the $d^7$ case. We parametrized  $\Delta_{\mathrm{trig}}=A \mathrm{cos}\phi, \lambda=A \mathrm{sin}\phi$, with $ A=50\ meV, \phi\in[0,\pi/2]$, while fixing the rest of the parameters to $U=6\ eV$, $U'=U-2J_H$, $J_H=1.0\ eV$, $\Delta_{\mathrm{oct}}=1.5\ eV$. The resulting total $J$ and decomposed total $S$ and $L$ moments are shown in Fig. \ref{fig:moments}.

In the limit of $\lambda \rightarrow 0$ ($\phi=0$) and $\Delta_{\mathrm{trig}} \ll \Delta_{\mathrm{oct}}$, a $S=3/2,L=0$ configuration is selected as the ground state for both cases, as expected.
On the other hand,  when the trigonal distortion is introduced,  the effective $\left\langle J^{\mathrm{eff}}_{z} \right\rangle$ moment saturates to a value of $J=1/2$. 
However, when $e_g$ orbitals are included, the total moment $\left\langle J^{\mathrm{total}}_z \right\rangle$ varies from 0.8 to 1.2 depending on the ratio of trigonal CFS and SOC, indicating an intermediate value between $S=3/2$ and $J=1/2$.

\section{Nuclear Refinements}\label{app:NuclearRefinements}

In addition to the neutron experiment on the MACS spectrometer, we acquired neutron diffraction data on the same sample using the BT1 powder diffractometer at the NCNR. We used 18.9 meV neutrons with 60' collimation before the monochromator and 20' collimation after the sample, measuring for 8.5 hours at 1.5 K, 6 hours at 4.8 K, and 6 hours at 28 K. These measurements cover a much larger $Q$ range than the MACS measurements with better Q-resolution, for a more complete determination of the nuclear structure.

\begin{figure}
\centering\includegraphics[scale=0.51]{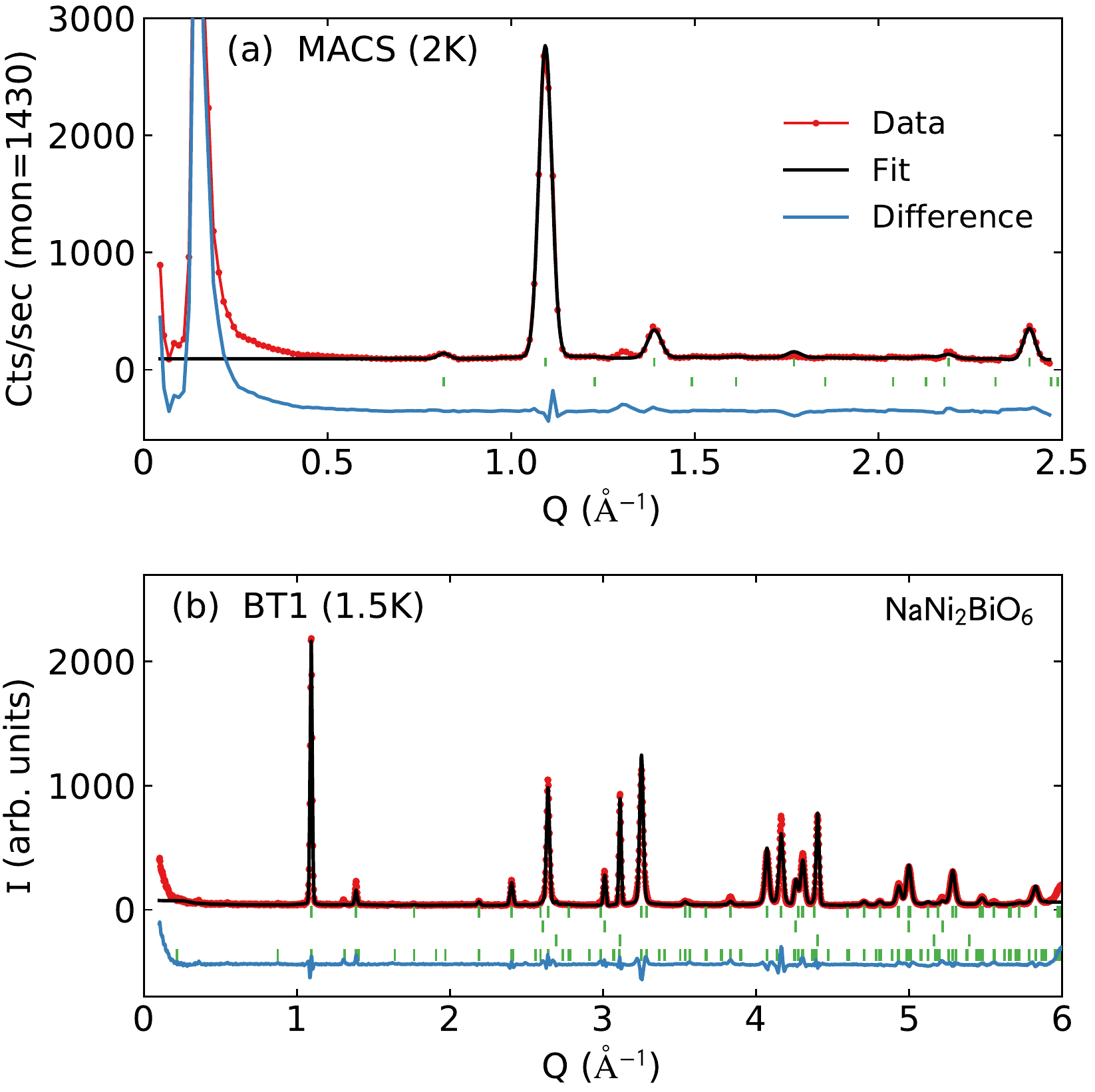}

\caption{Nuclear refinement of $\rm NaNi_2BiO_{6-\delta}$. (a) Refinement of unsubtracted MACS data at $2\>$K, which was used to define the peak widths and intensities of the magnetic peaks in the magnetic refinement. The vertical green lines show the peak locations from the nuclear and magnetic phases. Features for $Q<0.3 \AA^{-1}$ arise from instrumental direct beam backgrounds. (b) Refinement of BT1 data, showing fits to much higher $Q$ peaks. This refinement includes the nuclear phase, a NiO powder phase, aluminum from the sample can, and the magnetic phase.}

\label{flo:NuclearRefinement}
\end{figure}

Before refining the magnetic structure, we refined the nuclear structure using both the MACS and BT1 neutron diffraction data sets. The refinements are shown in Fig. \ref{flo:NuclearRefinement}. Both these data sets were taken below $T_{c2}$, and thus the refinements include the ${\bf q}=(\frac{1}{3},\> \frac{1}{3},\> 0.154)$ magnetic phase. The refinement in panel (a) includes a nuclear phase with Bragg peaks located as indicated by the upper vertical green lines, and a magnetic phase indicated by the lower row of vertical green lines. The refinement to the BT1 data set in panel (b) includes the nuclear phase (topmost vertical green lines), an additional NiO powder phase with 1.5\% of the refined intensity of the nuclear phase (second row of green lines), aluminum peaks from the sample can (third row of green lines), and the magnetic phase (fourth row of green lines).
The BT1 data do not show the magnetic structure as clearly as the MACS data---only the peak at 0.81 \AA$^{-1}$ is visible at 1.5 K---but the BT1 data include many more nuclear peaks and provide a better view of the nuclear structure.

\begin{table}
\caption{Refined nuclear positions for $\rm NaNi_2BiO_{6-\delta}$, $P \bar 31m$.}
\begin{ruledtabular}%
\begin{tabular}{cccccc}
atom type & label & $x$ & $y$ & $z$ & S.O.F. \tabularnewline
\hline 
Na & Na1 & 1/3 & 2/3 & 1/2 & 1  \tabularnewline
Ni & Ni1 & 1/3 & 2/3 & 0 & 1 \tabularnewline
Bi & Bi1 & 0 & 0 & 0 & 0.912 \tabularnewline
Bi & Bi2 & 0 & 0 & 0.114 & 0.080 \tabularnewline
O & O1 & 0.344(4) & 1.0 & 0.180(1) & 0.944 \tabularnewline
\end{tabular}\end{ruledtabular}
\label{flo:NucRefinementStructure}

\end{table}

The refined nuclear model, given in Table \ref{flo:NucRefinementStructure}, displayed in Fig. 1 of the main text, and described in detail in Ref \cite{Seibel2014}, provides a reasonable fit. However, some small peaks are not accounted for, most noticeably a weak Bragg peak at $Q = 1.3\>$\AA$^{-1}$. It is unclear what causes these deviations, whether there exists a nuclear supercell associated with oxygen vacancies or an additional phase in the sample.  Be that as it may, none of the unindexed peaks are temperature-dependent, which means that the magnetic signal from temperature subtraction is reliably from $\rm NaNi_2BiO_{6-\delta}$ alone. This magnetic signal can be indexed by a single ordering wave vector and fit to a consistent model based on the proposed  $\rm NaNi_2BiO_{6-\delta}$ chemical structure.




\section{Relating Neutron Bandwidth to Curie Temperature}\label{app:BandwidthToCW}

Based on a spin Hamiltonian 
$\mathcal{H}^{spin}=-\sum_{\langle ij \rangle}J_{ij}\mathbf{S}_{i}\cdot \mathbf{S}_{j}$ where $J$ represents bond energies, 
the Curie temperature (the temperature at which spontaneous magnetization occurs in the mean field approximation) is \cite{AshcroftAndMermin}
\begin{equation}
    \Theta_{CW}=zJ\frac{S\left(S+1\right)}{3k_{B}} .
    \label{eq:curieTemp}
\end{equation} 
This is the same Curie temperature which appears in the Curie-Weiss law $\chi=\frac{C}{T-\Theta_{CW}}$.

Meanwhile, the expression for a spin wave dispersion for collinear antiferromagnetic order is $\epsilon({\rm k}) =zJS\sqrt{(1+h_{a})^{2}-\gamma({\rm k})^{2}}$ (see for example Lovesey eq. 9.245 \cite{Lovesey}), where $\epsilon({\rm k})$ is the dispersion relation (determining the measured $\hbar \omega$ in spectroscopy), $z$ is the coordination number, $J$ is the exchange interaction, $h_{a}$ is the single ion anisotropy, and $\gamma({\rm k})=\frac{1}{z}\sum_{\delta}e^{i\mathbf{k}\cdot\mathbf{\delta}}$ where ${\bf \delta}$ lists the nearest neighbors. $\hbar \omega$ is maximal when $\gamma$ is minimal, which in the honeycomb lattice goes to zero when ${\bf k} = (0,\frac{4\pi a}{3\sqrt{3}})$, $a$ being the nearest neighbor distance. Thus, $\hbar\omega$ is maximal at
\begin{equation}
    \epsilon({\rm k})_{max}=zJS.
    \label{eq:swbandwidth}
\end{equation}

By combining eq. \ref{eq:curieTemp} with \ref{eq:swbandwidth}, we can solve for $\epsilon({\rm k})_{max}$ and relate the curie temperature to the neutron scattering bandwidth:
\begin{equation}
    \epsilon({\rm k})_{max}=\frac{3k_{B}}{(S+1)}\Theta_{CW}.
    \label{eq:PredictedBandwidth}
\end{equation}

\section{Magnetic Refinements}\label{app:GroupTheory}

\begin{figure}
\includegraphics[scale=0.51]{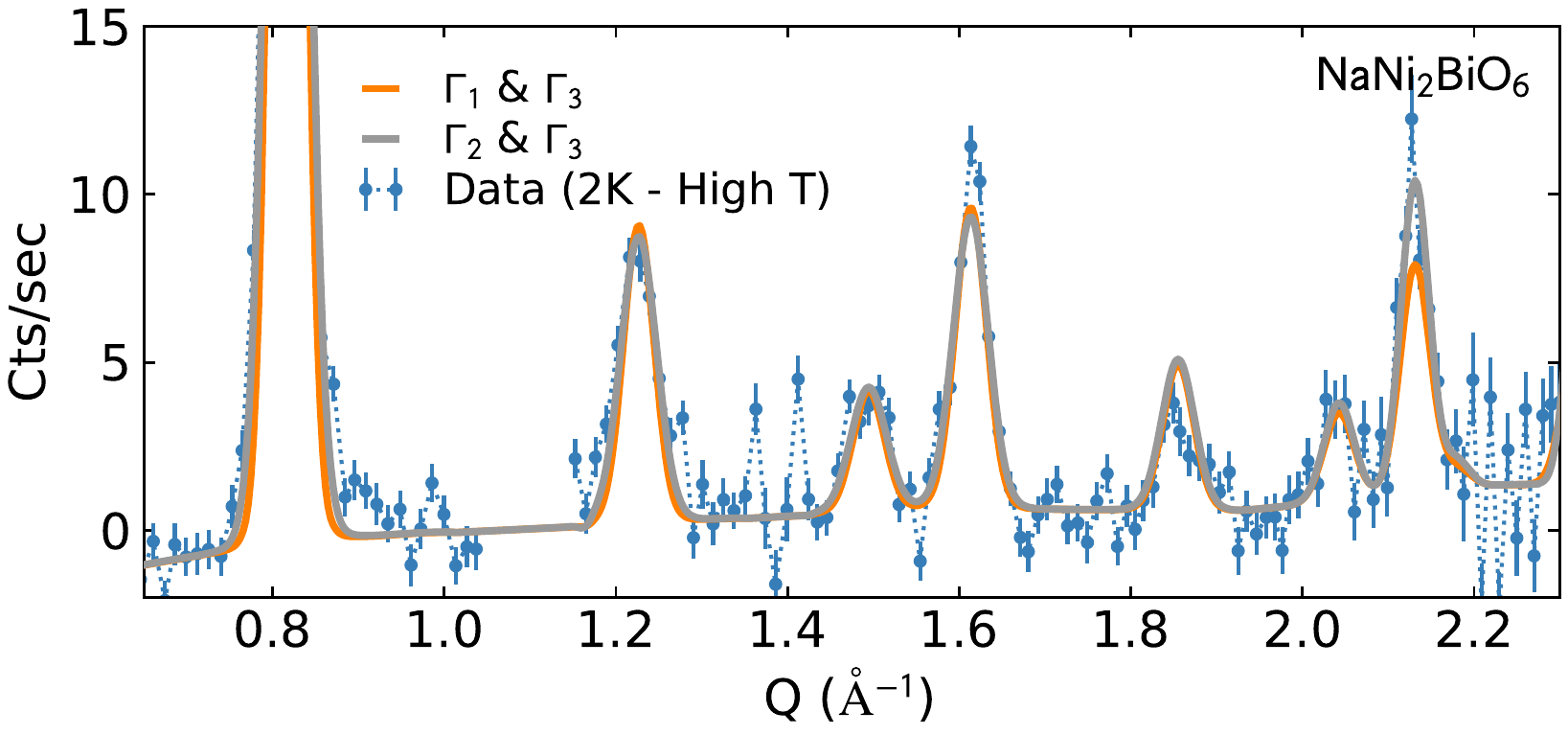}

\caption{Comparison between the magnetic refinements using $\Gamma_1 + \Gamma_3$ (grey line) and $\Gamma_2 + \Gamma_3$ (orange line). The differences between the two structures are too subtle to differentiate using the neutron diffraction data from our experiments.}

\label{flo:GammaComparison}
\end{figure}

\subsection{Irrep Decomposition}
Here we summarize our analysis to generate basis vectors of the $P \bar 31m$ space group with the ordering vector ${\bf q}=(\frac{1}{3},\> \frac{1}{3},\> 0.154)$.

Space group $P\overline{3}1m$ (also written $D_{3d}^{1}$ in Schoenflie notation) has 12 point symmetry operations. Half of them preserve ${\bf k}=(\frac{1}{3},\frac{1}{3},0.15)$ up to a reciprocal lattice vector, which is an ordering vector of type ${\bf k}_{10}$ in Kovalev's notation, yielding a group $G_k$ of the propagation wave vector with the following point operations:
$$
G_k= \{ h_i | t \} = 
\begin{cases}
h_{1} & (x,y,z)\\
h_{3} & (-y,x-y,z)\\
h_{5} & (-x+y,-y,z)\\
h_{20} & (y,x,z)\\
h_{22} & (-x,-x+y,z)\\
h_{24} & (x-y,-y,z)
\end{cases}.
$$
where the unit vectors (100), (010), and (001) are along the $a$, $b$, and $c$ axes respectively.
Generating the permutation, axial, and magnetic representations yields a character table in Table \ref{flo:characterTable}. In single valued representations, there are three irreducible representations listed in Kovalev's tables \cite{Kovalev}, shown in Table \ref{flo:Irreps}.

\begin{table}
\caption{Character table for $P \bar 31m$ with propagation vector ${\bf k}_{10}$.}
\begin{tabular}{c|c|c|c|c|c|c}
 & $h_{1}$ & $h_{3}$ & $h_{5}$ & $h_{20}$ & $h_{22}$ & $h_{24}$\tabularnewline
\hline 
$\chi_{axial}$ & 3 & 0 & 0 & -1 & -1 & -1\tabularnewline
$\chi_{perm}$ & 2 & 2 & 2 & 0 & 0 & 0\tabularnewline
$\chi_{mag}$ & 6 & 0 & 0 & 0 & 0 & 0\tabularnewline
\end{tabular}\label{flo:characterTable}
\end{table}

\begin{table}
\caption{Irreducible representations for $P \bar 31m$ with propagation vector ${\bf k}_{10}$. $\varepsilon=e^{i\frac{2\pi}{3}}=-0.5+\frac{\sqrt{3}}{2}i$.}
\begin{tabular}{c|cccccc}
${\bf k}_{10}$ & $h_{1}$ & $h_{3}$ & $h_{5}$ & $h_{20}$ & $h_{22}$ & $h_{24}$\tabularnewline
\hline 
$\Gamma_{1}$ & 1 & 1 & 1 & 1 & 1 & 1\tabularnewline
$\Gamma_{2}$ & 1 & 1 & 1 & -1 & -1 & -1\tabularnewline
$\Gamma_{3}$ & 
{\scriptsize
$\left(\begin{array}{cc}
1 & 0\\
0 & 1
\end{array}\right)$} & {\scriptsize$\left(\begin{array}{cc}
\varepsilon & 0\\
0 & \varepsilon^{2}
\end{array}\right)$} & {\scriptsize$\left(\begin{array}{cc}
\varepsilon^{2} & 0\\
0 & \varepsilon
\end{array}\right)$} & {\scriptsize$\left(\begin{array}{cc}
0 & 1\\
1 & 0
\end{array}\right)$} & {\scriptsize$\left(\begin{array}{cc}
0 & \varepsilon\\
\varepsilon^{2} & 0
\end{array}\right)$} & {\scriptsize$\left(\begin{array}{cc}
0 & \varepsilon^{2}\\
\varepsilon & 0
\end{array}\right)$}\tabularnewline
\end{tabular}\label{flo:Irreps}
\end{table}

Given their dimensionality, $\Gamma_1$ and $\Gamma_2$ have one basis vector and $\Gamma_3$ has two basis vectors.
To find the basis vectors, we project onto the test functions: $\phi_{1}=(1\,0\,0)$,
$\phi_{2}=(0\,1\,0)$, $\phi_{3}=(0\,0\,1)$. Using the projection equation \cite{Izyumov1979}
\begin{equation}
{\psi}_{\alpha\nu}^{\lambda}=\sum_{g\in G_{k}}D_{\nu}^{\lambda}*(g)\sum_{i}e^{-i{\bf  q}\cdot{\bf a}_{gi}}\delta_{i,gi}\Gamma_{axial}^{g}{\phi_{\alpha}}
\end{equation}
 we have, throwing away all the zero pairs of basis vectors, the set of basis vectors listed in Table \ref{flo:OriginalIrrRepTable}.

As is immediately clear, this procedure yields more than two basis vectors for $\Gamma_3$ which therefore cannot be orthogonal. Thus, we must combine them into two pairs of linear combintions. Two of the basis vectors describe one triangular Ni sub-lattice site and two describe the other equivalent Ni lattice, the two together forming the honeycomb structure. The sets of basis vectors are identical except for a sign change, but they describe the two lattice sites separately. 
Linear combinations that link the two sublattices take the form $\psi_{net} = x\psi_4 + (1 - x)\psi_6$ (where $0 \leq x \leq 1$), and the diffraction pattern is independent of $x$. 
The existence of two phase transitions, however, requires that site equivalency be enforced. A value of $x=1/2$ would result in $Pm$ symmetry, but any value other than $1/2$ would result in $P1$ symmetry (only the identity operation)---which has no symmetry elements beyond translations. This precludes a second order phase transition to a lower-symmetry state that does not modify the magnetic wave vector. We do have an additional phase transition at $T_{c2}$, so a point group symmetry must remain for $T_{c2}<T<T_{c1}$. Therefore, we neglect the possibility of spontaneous sublattice symmetry breaking and set $x=1/2$. This results in the basis vectors listed in Table I of the main text. The in-plane 120\textdegree exchange bond-dependent correlations are described by the one-dimensional irrep $\Gamma_2$ which is not subject to these considerations.

\begin{table}
\caption{Original irreducible representations and associated basis vectors for space
group $P \bar 31m$ and ordering wave vector ${\bf q} =(\frac{1}{3},\frac{1}{3},0.154)$.}
\begin{ruledtabular}%
\begin{tabular}{ccccc}
IRs & $\psi_{\nu}$ & component & Ni1 & Ni2  \tabularnewline
\hline 
$\Gamma_{1}$  & $\psi_{1}$ & Real & (1.5 0 0) & (0 -1.5 0)   \tabularnewline
 &  & Imaginary & ($-\frac{\sqrt{3}}{2}$ $-\sqrt{3}$ 0) & ($\sqrt{3}$ $\frac{\sqrt{3}}{2}$ 0)   \tabularnewline
$\Gamma_{2}$  & $\psi_{2}$ & Real & (1.5 0 0) & (0 1.5 0)  \tabularnewline
 &  & Imaginary & ($-\frac{\sqrt{3}}{2}$ -$\sqrt{3}$ 0) & (-$\sqrt{3}$ $-\frac{\sqrt{3}}{2}$ 0) \tabularnewline
$\Gamma_{3}$  & $\psi_{3}$ & Real & (0 0 0) & (0 -1.5 0)  \tabularnewline
 &  & Imaginary & (0 0 0) & (-$\sqrt{3}$ $-\frac{\sqrt{3}}{2}$ 0) \tabularnewline
 & $\psi_{4}$ & Real & (0 0 3) & (0 0 0)  \tabularnewline
 &  & Imaginary & (0 0 0) & (0 0 0)  \tabularnewline
 & $\psi_{5}$ & Real & (1.5 0 0) & (0 0 0)  \tabularnewline
 &  & Imaginary & ($\frac{\sqrt{3}}{2}$ $\sqrt{3}$ 0) & (0 0 0) \tabularnewline
 & $\psi_{6}$ & Real & (0 0 0) & (0 0 -3)  \tabularnewline
 &  & Imaginary & (0 0 0) & (0 0 0)  \tabularnewline
\end{tabular}\end{ruledtabular}
\label{flo:OriginalIrrRepTable}

\end{table}

\subsection{In-plane Structure and Symmetry}

As noted in the text, there are two combinations of irreducible representations that fit the low temperature phase of $\rm NaNi_2BiO_{6-\delta}$: $\Gamma_1 + \Gamma_3$ (with antiferromagnetic 120\textdegree exchange in-plane correlations) and $\Gamma_2 + \Gamma_3$ (with ferromagnetic 120\textdegree exchange in-plane correlations). The two different predicted diffraction patterns are shown in Fig. \ref{flo:GammaComparison}, with peak widths fit to the magnetic peaks. There are subtle differences between the patterns, but the differences are so small that we are unable to distinguish between them with neutron diffraction. Therefore, we look to symmetry considerations to determine the correct ground state.

Second order phase transitions are directly associated with symmetry breaking. This means that for in-plane spin ordering to account for the phase transition that we observe at $T_{c2}$, the in-plane spin structure must break a symmetry operation of the intermediate temperature phase. This allows us to identify the in-plane spin order for $T<T_{c2}$.

The $\psi_4$ structure of the intermediate temperature ($T_{c2} < T < T_{c1}$) phase, when site-equivalency is enforced, has only two valid symmetry operations from the group of the propagation vector $G_k$: $h_1$ (identity) and $h_{20}$ (reflection about [110]). 
Meanwhile, $\Gamma_1$ preserves all $G_k$ symmetries ($h_{1}$, $h_{3}$, $h_{5}$, $h_{20}$, $h_{22}$, and $h_{24}$), $\Gamma_2$ preserves all the 3-fold rotation symmetries but no mirror planes ($h_{1}$, $h_{3}$, and $h_{5}$), and the $\Gamma_3$ in-plane structure $\psi_3$ preserves only the identity and the (110) mirror plane ($h_{1}$ and  $h_{20}$). 
A $\Gamma_1 + \Gamma_3$ structure would have $h_{1}$ and  $h_{20}$ symmetry, resulting in no broken symmetries. A $\Gamma_2 + \Gamma_3$ structure would have only $h_{1}$ symmetry, resulting in a broken symmetry. A $\psi_3 + \psi_4$ structure would have $h_{1}$ and  $h_{20}$ symmetry, resulting in no broken symmetries. The only in-plane structure that breaks a symmetry of the intermediate temperature phase is $\Gamma_2$. This means the $T<T_{c2}$ magnetic structure of  $\rm NaNi_2BiO_{6-\delta}$ must form a reducible representation of $G_k$ based on irreps $\Gamma_2 + \Gamma_3$. The corresponding spin structure is depicted in Fig. 7(c) of the main text.

\subsection{Correlation Length}\label{app:CorrelationLength}

\begin{figure}
\includegraphics[scale=0.51]{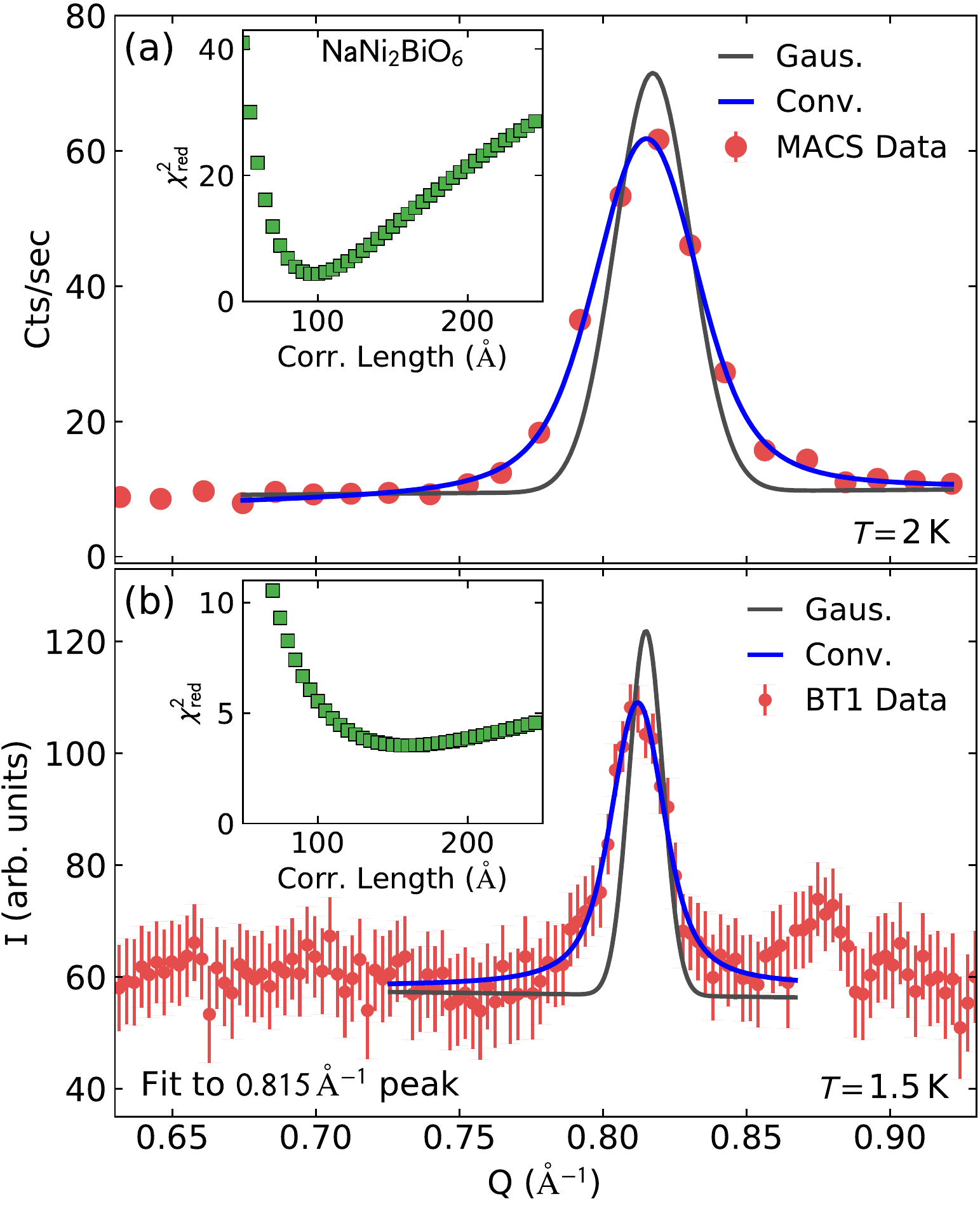}

\caption{Profile fits to the  0.81 \AA$^{-1}$ magnetic Bragg peak, with the data in red, the original Gaussian fit with the with the width defined by the nuclear peaks in black, and the best fit convoluted profile in blue. (a) shows fits to MACS data, and (b) shows fits to BT1 data. The insets show reduced $\chi^2$ of a convoluted profile fit vs magnetic correlation length.}

\label{flo:CorrelationLength}
\end{figure}

The peak widths of the nuclear Bragg peaks in Fig. \ref{flo:NuclearRefinement} are smaller than the magnetic Bragg peaks widths in the temperature-subtracted data. This indicates the magnetic correlation length is less than that of the underlying crystal structure. To determine  the magnetic correlation length, we fit the the strongest magnetic Bragg peak (at 0.81 \AA$^{-1}$) with a convolution of a Gaussian (with peak width defined by the nuclear phase) and a Lorentzian profile, as shown in Fig. \ref{flo:CorrelationLength}. The inverse of the Lorentzian HWHM is the magnetic correlation length, which has a best fit value $97 \pm 7$ \AA, indicated by the minimum in reduced $\chi^2$ of the convoluted profile fit in Fig. \ref{flo:CorrelationLength}.

The 0.81 \AA$^{-1}$ peak is at $(000) + {\bf q}$ peak, where ${\bf q}$ is the magnetic propagation vector $(\frac{1}{3},\> \frac{1}{3},\> 0.154 \pm 0.011)$.
The remainder of the magnetic Bragg peaks are much weaker and the fits are consequently less reliable, but the results are consistent: fitting the MACS data for 1.22 \AA$^{-1}$ [$(001) - {\bf q}$], 1.61 \AA$^{-1}$ [$(110)  - {\bf q}$], and 2.12 \AA$^{-1}$ [$(100)  + {\bf q}$] peaks simultaneously yielded a correlation length of $116 \pm 30$ \AA, which agrees with the fit of the 0.81 \AA$^{-1}$ peak to within uncertainty. These peaks are  27\textdegree,  85\textdegree, and  86\textdegree\, from the $c$ axis respectively, which means the first is associated mostly with $c$ axis correlations and the last two are associated with in-plane correlations. If we treat the $(001) - {\bf q}$ peak (mostly along the $c$ axis) separate from $(100)  + {\bf q}$ and $(110)  - {\bf q}$ (mostly in-plane), we find a correlation length of $57 \pm 9$ \AA\, for $(001) - {\bf q}$ and a correlation length of $160 \pm 70$ \AA\, for the $(100)  + {\bf q}$ and $(110)  - {\bf q}$ peaks. This indicates a correlation length three times smaller along the $c$-axis than in the plane, consistent with a quasi-2D magnetic material.

We carried out the same analysis on the 0.81 \AA$^{-1}$ peak from the BT1 data in Fig. \ref{flo:CorrelationLength}(b), and found a correlation length of $152 \pm 16$ \AA. We consider this value to be more reliable than the MACS data because the BT1 nuclear peak width is defined by many peaks and is well constrained, but the nuclear peak width for the MACS data is defined only by three peaks (see Fig. \ref{flo:NuclearRefinement}) and is underconstrained.

\section{Luttinger Tisza Analysis}\label{app:LuttingerTisza}

As noted in the text, the magnetic ordering wave vector of ${\bf q}=(\frac{1}{3},\> \frac{1}{3},\> 0.154 \pm 11)$ is unusual for the honeycomb lattice, both because of the $c$ axis modulation and the in-plane $(\frac{1}{3},\> \frac{1}{3})$ order. To explore whether such a magnetic ordering wave vector can be stabilized by Heisenberg interactions at the mean-field level, we used Luttinger-Tisza theory \cite{LuttingerTisza}. While this method has its limitations and does not consider emergent interactions resulting from thermal or quantum fluctuations, it does give a basic picture of what orders are readily stabilized.

We began with the Fourier transform of the Heisenberg Hamiltonian for helical order:
\begin{equation}
    \mathcal{H}(\mathbf{q})=\frac{1}{2}\sum_{\nu\mu}J_{\nu\mu}(\mathbf{q})\mathbf{S}_{q\nu}\cdot \mathbf{S}_{q\mu}^{*}
\end{equation}
where $\nu$ and $\mu$ sum over sites within the paramagnetic unit cell and
\begin{equation}
    J_{\nu\mu}(\mathbf{q})=\sum_{\Delta\mathbf{R}}J_{\nu\mu}(\Delta\mathbf{R})e^{i\mathbf{q}\cdot \Delta\mathbf{R}_{\nu\mu}}.
\end{equation}
With two nickel sites per unit cell, the Hamiltonian can be written as a $2\times2$ matrix whose smallest eigenvalue is the minimum energy. Although $J_{\nu\mu}$ is complex, when summing over all atoms in the unit cell the eigenvalues of the matrix are always real.
With this equation, one can find the ordering wave vector $\bf q$ which minimizes $\mathcal{H}^{spin}$ for a given set of $J(\Delta\mathbf{R})$---i.e., we identify the the magnetic wave vector stabilized by a given set of interactions.

To search for a set of exchange constants which stabilize $q=(\frac{1}{3},\> \frac{1}{3},\> \eta)$, we systematically defined a series of exchange constants and found the wave vector  $\bf q$ minimizing $\mathcal{H}(\mathbf{q})^{spin}$. We used a L-BFGS-B minimization routine \cite{BFGS}, always with $(\frac{1}{3},\> \frac{1}{3}, 0.154)$ as the starting $\bf q$.

\begin{figure}
\centering\includegraphics[scale=0.09]{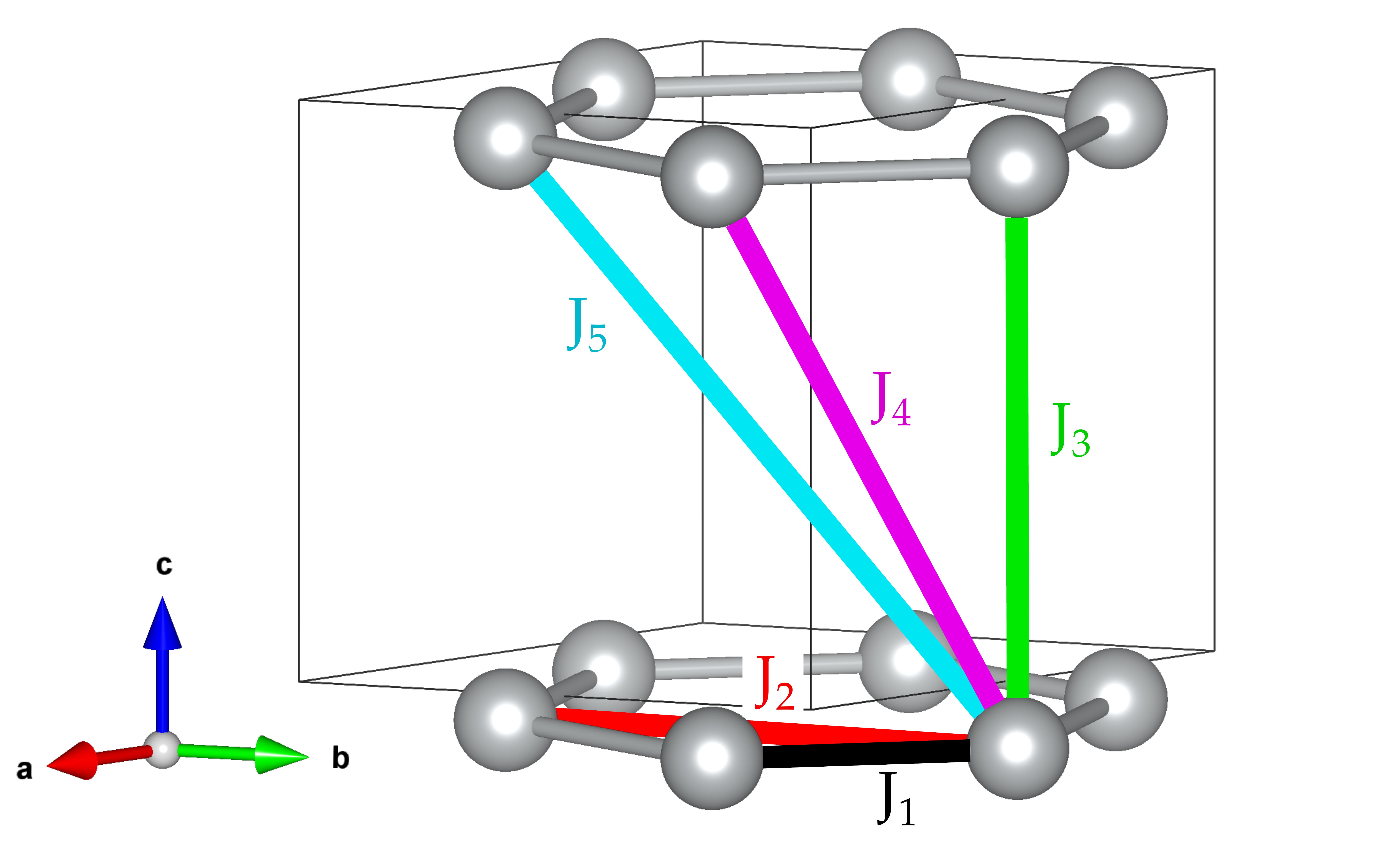}

\caption{Exchange constants considered in the Luttinger Tisza analysis of $\rm NaNi_2BiO_{6-\delta}$. $J_1$ and $J_2$ are in-plane interactions while $J_3$, $J_4$, and $J_5$ are out-of-plane interactions.}

\label{flo:ExchangeConstants}
\end{figure}

In our analysis we considered five exchange  interactions: two in-plane, and three out-of plane  interactions (Fig. \ref{flo:ExchangeConstants}). The super-exchange paths of all three out of plane interactions involve the same number of atoms: Ni-O-Na-O-Ni (Fig. \ref{flo:ExchangeConstants}), which indicates their strength could be comparable. Including both Ni sites in the Hamiltonian ensured that $\mathcal{H}^{spin}$ is always real.
We set $J_1=1$ (nearest neighbor exchange), and let the other interactions vary from -1 to 1, and $J_2$ from -2 to 2.

As a rough cross-check, we compared our Luttinger-Tisza results to more sophisticated calculations of the honeycomb phase diagram \cite{Albuquerque2011, Li2012}. In our calculations the transition from (0,0) to (1/2, 1/2) occurs at $J_2/J_1 = 0.17$ when all other interactions are zero. In refs. \cite{Albuquerque2011, Li2012}, the transition is closer to $J_2/J_1 = 0.2$, though there is an intermediate disordered phase in between that does not appear in the Luttinger-Tisza analysis.

\begin{figure}
\centering\includegraphics[scale=0.1]{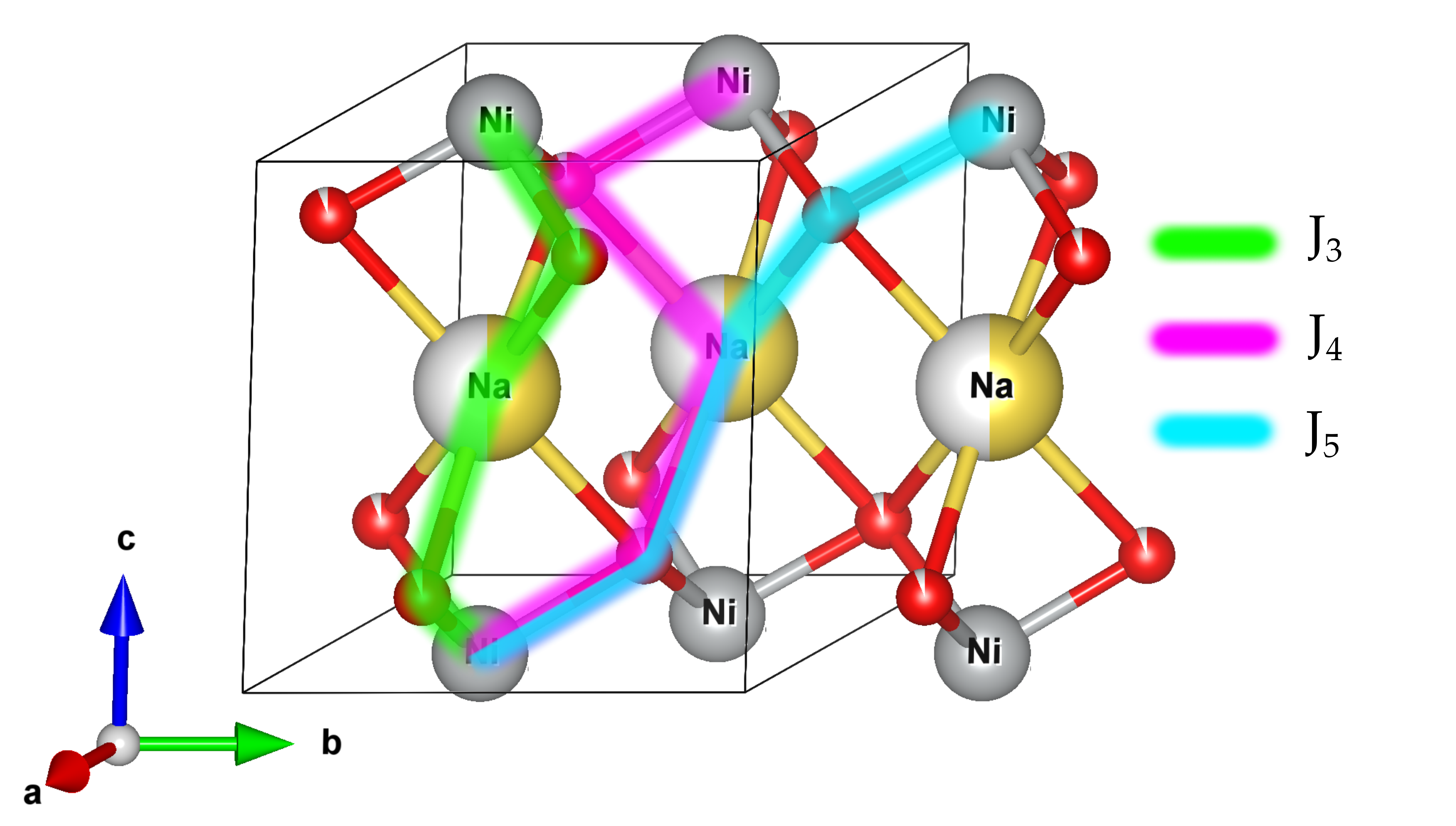}

\caption{Out-of-plane exchange pathways in $\rm NaNi_2BiO_{6-\delta}$. The first (green), second (purple), and third (blue) nearest inter-honeycomb-plane exchange pathways all follow a Ni-O-Na-O-Ni pathway, which means they could be of comparable strength.}

\label{flo:InterlayerCoupling}
\end{figure}

\begin{figure*}
\centering\includegraphics[scale=0.5]{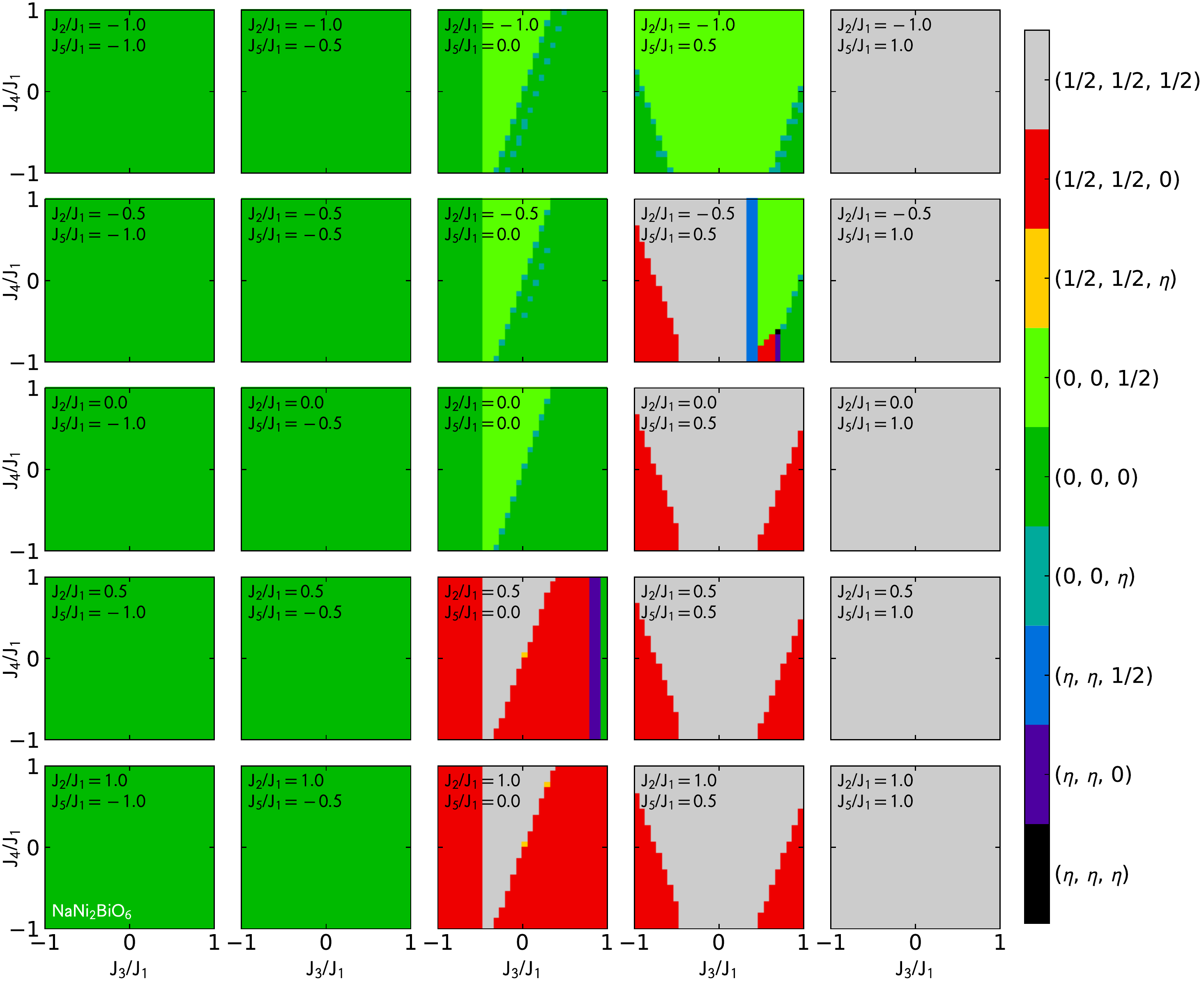}

\caption{Ordering wave vectors stabilized by isotropic exchange interactions $J_2$, $J_3$, $J_4$, and $J_5$ (defined in Fig. \ref{flo:ExchangeConstants}) relative to $J_1$, as calculated by Luttinger Tisza theory. Each panel shows a range of $J_3/J_1$ and $J_4/J_1$ values for specific values of $J_2/J_1$ and $J_5/J_1$. The ordering vector is indicated by the colorscale to the right. $\eta$ refers to an  wavelength other than 1, 1/2, or 1/3, which only appears on the boundaries between other phases.}

\label{flo:LuttingerTiszaResult}
\end{figure*}

\begin{figure*}
\centering\includegraphics[scale=0.57]{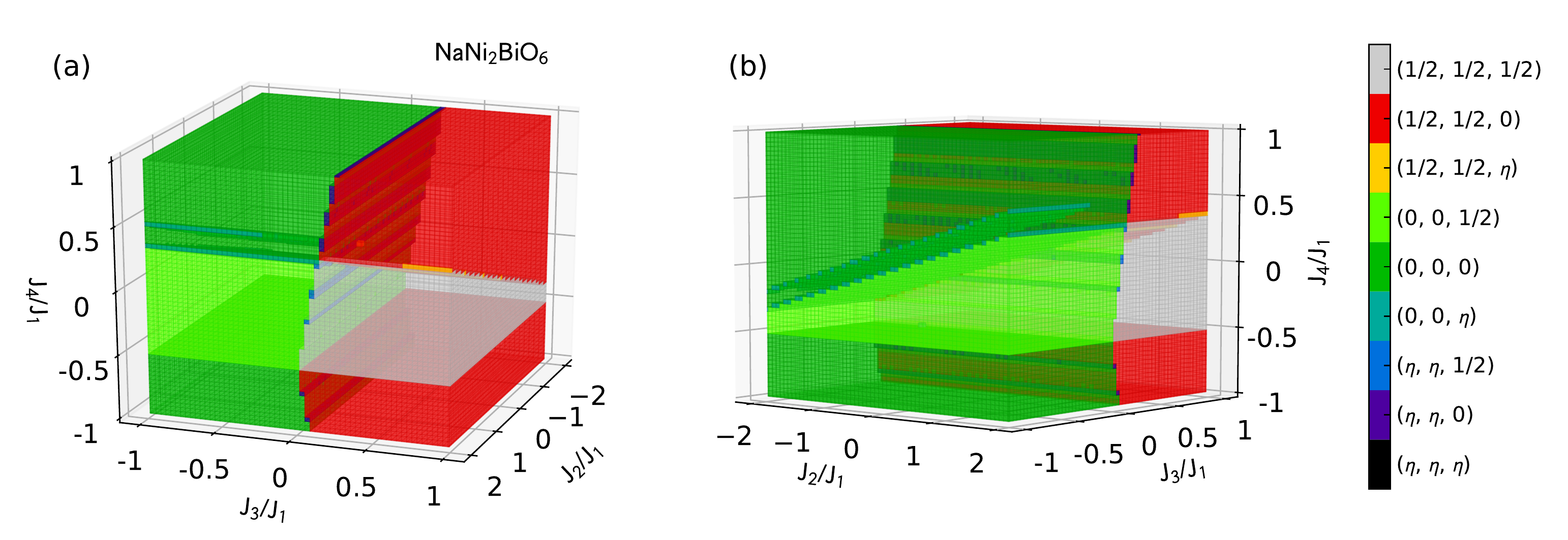}

\caption{3D plot of ordering wave vectors stabilized by isotropic exchange interactions $J_2$, $J_3$, and $J_4$ relative to $J_1$ with $J_5 = 0$, as calculated by Luttinger Tisza theory. (a) and (b) show two views of the same data set. The ordering vector is indicated by the colorscale to the right. Note that incommensuarate order (ordering vectors with $\eta$) only appears on the boundaries between other phases.}

\label{flo:LuttingerTiszaResult3d}
\end{figure*}

A selection of results of this analysis are shown in Figs. \ref{flo:LuttingerTiszaResult} and \ref{flo:LuttingerTiszaResult3d}. Most of the exchange parameter space considered stabilizes commensurate order, but never $(\frac{1}{3},\> \frac{1}{3})$ in-plane order. The boundaries between phases [for example between $(\frac{1}{2},\> \frac{1}{2},\> 0)$ and $(\frac{1}{2},\> \frac{1}{2},\> \frac{1}{2})$] sometimes show incommensurability, but "zooming in" and increasing the resolution of the parameter search shows the incommensurate regions exist only on the boundaries. No finite regions of parameter space stabilize $q=(\frac{1}{3},\> \frac{1}{3},\> \eta)$ order, much less the observed $\eta \approx 1/6$.
The failure to account for the observed magnetic order with a Heisenberg model suggests that the exchange Hamiltonian $\rm NaNi_2BiO_{6-\delta}$ is not  isotropic and this concords with the analysis presented in the main text that the counterrotating state is not favored by conventional bond independent exchange interactions. 

\section{Anisotropic exchange and $c$-axis modulation}\label{app:Incommensurability}

\begin{figure}
\centering\includegraphics[scale=0.5]{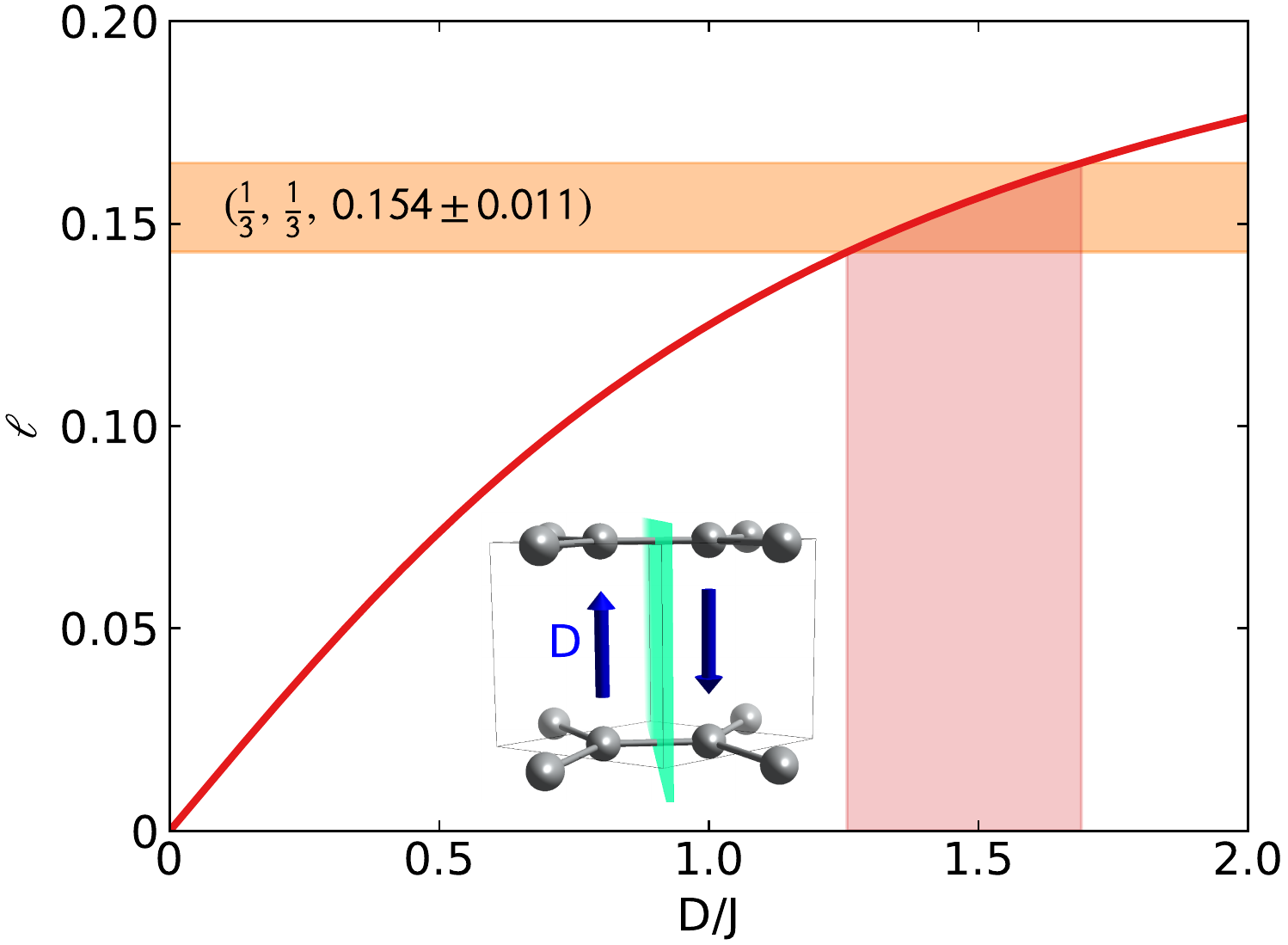}

\caption{Magnetic propogation vector ${\bf q}=(\frac{1}{3},\> \frac{1}{3},\> \ell)$ stabilized by various values of $D/J$: ratio of inter-plane DM exchange to inter-plane Heisenberg exchange. The observed $\ell = 0.154$ order means that $D/J = 1.5(2)$. The inset shows the DM vectors as constrained by the symmetry of the lattice.}

\label{flo:DMexchange}
\end{figure}

The $q_c = 0.154$ long-wavelength order along the $c$ axis can be explained by invoking anisotropic exchanges. 
The Dzyaloshinskii-Moriya (DM) exchange ${\bf D} \cdot ({\bf S} \times {\bf S})$ appears when there is not inversion symmetry at the midpoint between sites \cite{Moriya}. This is the case for inter-plane exchange on the  $\rm NaNi_2BiO_{6-\delta}$ lattice, because the honeycomb lattice itself lacks inversion symmetry at the magnetic sites. The three-fold rotation symmetry about this bond further constrains ${\bf D}$ to be along the $c$-axis, and the mirror symmetry between the two Ni sites inverts the DM vector between the two sites as shown in Fig. \ref{flo:DMexchange}. So for $\rm NaNi_2BiO_{6-\delta}$, we can use symmetry to identify the direction  of ${\bf D}$ precisely.

This DM exchange, when in competition with an inter-plane ferromagnetic Heisenberg exchange, produces a long-wavelength (generally incommensurate) spiral order along the $c$-axis. It also produces counter-rotating spiral spins on the two different Ni sites (due to the flipped DM vector on the different sites), just as observed in the neutron diffraction refinements. If the DM vector is around 1.5 times as strong as the inter-plane Heisenberg exchange (which we expect to be around 0.1 meV from comparisons with $\rm Ni_2O_3$), we produce exactly the observed $\ell \approx 1/6$ ordering vector with the correct in-plane structure.
However, the fact that the $c$-axis wave vector is the same in the intermediate-temperature collinear phase suggests that something beyond the DM interaction is at play because the DM exchange only acts upon in-plane moments.

Another possibility is the biquadratic exchange $J_{bq}({\bf S}_i \cdot {\bf S}_j)^2$, which can produce long-wavelength order when competing with a bilinear Heisenberg exchange $J_{bl}({\bf S}_i \cdot {\bf S}_j)$ of the opposite sign. Specifically, the wave vector is
\begin{equation}
    \ell = \frac{2\pi}{\cos^{-1}(\frac{-J_{bl}}{2 J_{bq}})},
\end{equation}
but this requires that $ |J_{bq}| > 0.5 |J_{bl}|$, which may or may not be realistic for $\rm NaNi_2BiO_{6-\delta}$.



\begin{thebibliography}{68}
\bibitem [{NIS()}]{NIST_disclaimer}%
  \BibitemOpen
  \href@noop {} {}\bibinfo {note} {Certain commercial instruments are
  identified in this paper to foster understanding. Such identification does
  not imply recommendation or endorsement by the National Institute of
  Standards and Technology, nor does it imply that the instruments identified
  are necessarily the best available for the purpose.}\BibitemShut {Stop}%
\end{thebibliography}
\end{document}